\documentclass[%
 reprint,
 amsmath,amssymb,superscriptaddress,
 aps,
 prd,
 floatfix
]{revtex4-1}

\usepackage{graphicx}
\usepackage{float}
\usepackage{physics}
\usepackage[dvipsnames]{xcolor}
\usepackage[shortlabels]{enumitem}

\usepackage[normalem]{ulem}

\newcommand{\avr}[1]{\left\langle{#1}\right\rangle}

\allowdisplaybreaks
\begin{document}

\title{QCD deconfinement transition line up to $\mu_B=400$~MeV
from finite volume lattice simulations}

\author{Szabolcs Bors\'anyi}
\affiliation{Department of Physics, Wuppertal University, Gaussstr.  20, D-42119, Wuppertal, Germany}

\author{Zolt\'an Fodor}
\affiliation{Department of Physics, Wuppertal University, Gaussstr.  20, D-42119, Wuppertal, Germany}
\affiliation{Pennsylvania State University, Department of Physics, State College, PA 16801, USA}
\affiliation{Institute  for Theoretical Physics, ELTE E\"otv\"os Lor\' and University, P\'azm\'any P. s\'et\'any 1/A, H-1117 Budapest, Hungary}
\affiliation{J\"ulich Supercomputing Centre, Forschungszentrum J\"ulich, D-52425 J\"ulich, Germany}

\author{Jana N. Guenther}
\affiliation{Department of Physics, Wuppertal University, Gaussstr.  20, D-42119, Wuppertal, Germany}

\author{Paolo Parotto}
\affiliation{Dipartimento di Fisica, Universit\`a di Torino and INFN Torino, Via P. Giuria 1, I-10125 Torino, Italy}

\author{Attila P\'asztor}
\affiliation{Institute  for Theoretical Physics, ELTE E\"otv\"os Lor\' and University, P\'azm\'any P. s\'et\'any 1/A, H-1117 Budapest, Hungary}

\author{Ludovica Pirelli}
\affiliation{Department of Physics, Wuppertal University, Gaussstr.  20, D-42119, Wuppertal, Germany}

\author{K\'alm\'an K. Szab\'o}
\affiliation{Department of Physics, Wuppertal University, Gaussstr.  20, D-42119, Wuppertal, Germany}
\affiliation{J\"ulich Supercomputing Centre, Forschungszentrum J\"ulich, D-52425 J\"ulich, Germany}

\author{Chik Him Wong}
\affiliation{Department of Physics, Wuppertal University, Gaussstr.  20, D-42119, Wuppertal, Germany}


\date{\today}

\begin{abstract}

The QCD cross-over line in the temperature ($T$) -- baryo-chemical potential ($\mu_B$) plane
has been computed by several lattice groups
by calculating the
chiral order parameter and its susceptibility at
finite values of $\mu_B$. 
In this work we focus on the deconfinement aspect of the transition between hadronic and Quark Gluon Plasma (QGP) phases.
We define the deconfinement temperature as the peak position
of the static quark entropy ($S_Q(T,\mu_B)$) in $T$, 
which is based on the renormalized Polyakov loop. 
We extrapolate $S_Q(T,\mu_B)$ based on high statistics
finite temperature ensembles on a $16^3\times 8$ lattice to finite density
by means of a Taylor expansion to eighth order in $\mu_B$ (NNNLO)
along the strangeness neutral line.
For the simulations the 4HEX staggered action was used with 2+1 flavors at physical quark masses.
In this setup the phase diagram can be drawn up to unprecedentedly high chemical potentials. 
Our results for the deconfinement 
temperature are in rough agreement with phenomenological estimates of the freeze-out curve in relativistic heavy ion collisions.
In addition, 
we study the width of the deconfinement crossover. 
We show
that up to $\mu_B \approx 400$~MeV, the deconfinement
transition gets broader at higher densities, disfavoring the existence of a deconfinement critical endpoint
in this range.
Finally, we examine the transition line without the strangeness neutrality condition and 
observe a hint for the narrowing of the crossover towards large $\mu_B$.
\end{abstract}

\maketitle

\section{Introduction}

In the last decades, a large body of evidence has been gathered on the crossover 
transition between the hadronic 
and Quark Gluon Plasma (QGP) phases of QCD 
at zero net baryon density, both from  theory~\cite{Aoki:2006we, Borsanyi:2010bp, HotQCD:2019xnw, Bazavov:2011nk, Kotov:2021rah, Cuteri:2021ikv} 
and the phenomenology of  
heavy ion collision experiments~\cite{Pratt:2015zsa, Pang:2016vdc, STAR:2021rls}. Theoretically, there are
two distinct aspects of the QCD transition. One is
chiral symmetry restoration and the other is deconfinement. 

Chiral symmetry restoration is defined in the limit of zero light quark masses (the chiral limit), when chiral 
symmetry becomes exact. In this limit, it is generally expected that a genuine phase transition will take place, instead of a crossover, with the order of the transition depending on the number of quark flavours~\cite{Pisarski:1983ms, Pelissetto:2013hqa, Cuteri:2021ikv, Fejos:2024bgl}. The most relevant for phenomenology is the two-flavor chiral limit, where a second order transition in the $O(4)$ universality class is expected.
Chiral symmetry restoration is
studied mostly through the mass derivatives of the 
free energy: the chiral condensate and the 
(full or disconnected) chiral susceptibilities. Lattice studies have
determined the cross-over temperature in the case of physical quark
masses by locating the peak of these susceptibilities \cite{Aoki:2006br,Aoki:2009sc,Borsanyi:2010bp,Bazavov:2011nk}.
The behavior of these observables closer to the chiral limit was also 
studied on the lattice 
\cite{Ejiri:2009ac,Ding:2019prx,Ding:2024sux}.

\begin{figure*}[t]
\centerline{
\includegraphics[width=0.5\textwidth]{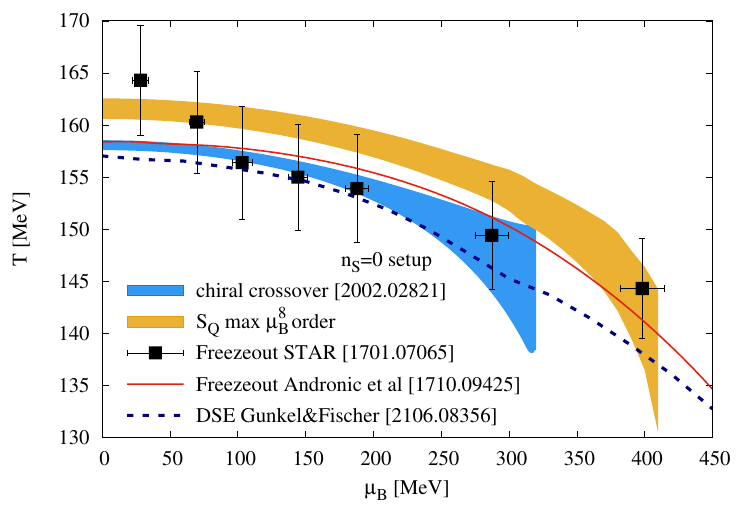}
\includegraphics[width=0.5\textwidth]{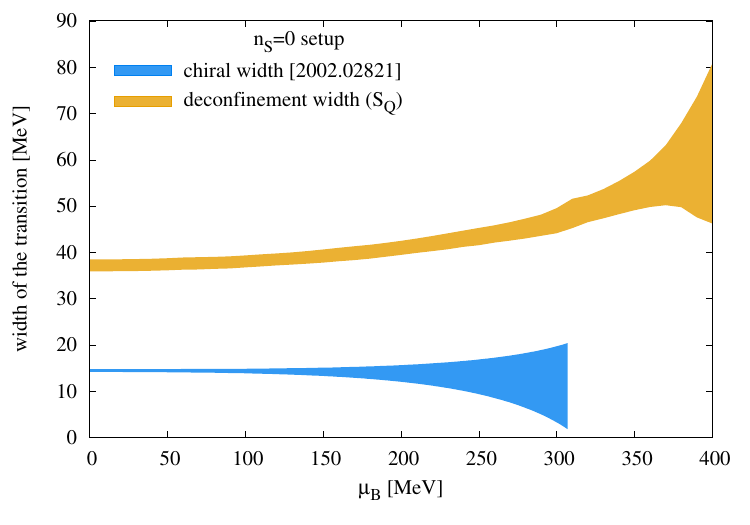}
}
\caption{
Left: the deconfinement line (orange) defined as the peak position of the
static quark entropy ($S_Q$) at fixed $\mu_B$ as a function of $\mu_B$.
We add the chiral transition line from lattice QCD \cite{Borsanyi:2020fev} in the
phase diagram,  defined from the peak of the full chiral susceptibility, as well as
the corresponding result based on Dyson-Schwinger equations \cite{Gunkel:2021oya}  
(the latter used the $\mu_S=\mu_B/3$ scheme as an approximation to $n_S=0$).
We also include the chemical freeze-out parameters \cite{STAR:2017sal,Andronic:2017pug}.
Right: the width associated to the deconfinement (orange)
and chiral (blue) transitions: the two differ already at $\mu_B=0$. With 
increasing $\mu_B$, while the chiral width is mostly constant, the 
deconfinement width grows. 
\label{fig:phase_diagram}
}
\end{figure*}

Deconfinement, on 
the other hand, is 
most cleanly defined in the limit of infinite quark masses, where center symmetry becomes exact. 
In this quenched limit there is a weak first order transition between the confined and deconfined gluon plasma phases (for a recent high precision study, see Ref.~\cite{Borsanyi:2022xml}).

An approximate order parameter for the deconfinement transition is the Polyakov loop, which probes the properties of infinite mass static test quarks in a QCD medium \cite{McLerran:1981pb}. Its use is not limited to infinite quark masses: it has proven useful to study the nature of the transition with heavy quarks \cite{Cuteri:2020yke}, even down to physical masses \cite{Aoki:2006br,Bazavov:2009zn,Bazavov:2013yv,Borsanyi:2015yka}.
It was also studied for smaller-than-physical quark masses, where it was observed that a rapid rise of the Polyakov loop still accompanies the chiral
transition
\cite{Clarke:2020htu}.

Several observables can be derived from the Polyakov loop to study confinement.
A good review of these is Ref.~\cite{Bazavov:2016uvm}, where it is demonstrated
that the static quark entropy has some favorable properties to study 
deconfinement numerically on the lattice.
Furthermore, it also has a peak position that is very close
to the value of the chiral crossover temperature.

While the crossover nature of the transition is very well established at zero baryon density (or baryochemical potential $\mu_B$), it is predicted by model and functional calculations that the transition line in the $T - \mu_B$ plane turns into a first order transition at higher baryon densities at a critical endpoint~\cite{Kovacs:2016juc,Gao:2020qsj,Fu:2021oaw,Isserstedt:2019pgx,Gunkel:2021oya,Critelli:2017oub,Hippert:2023bel}.
The experimental discovery of the QCD critical 
endpoint is a major goal of relativistic heavy 
ion collision experiments~\cite{STAR:2020tga}. 

However, at this point in time very little is known for certain
about the QCD phase diagram. One example is the curvature of the 
crossover temperature in $\mu_B$. Defined via observables 
related to chiral symmetry restoration, it is currently very 
well established~\cite{Bonati:2018nut, 
HotQCD:2018pds, Borsanyi:2020fev,Ding:2024sux}. The chiral transition line $T_c(\mu_B)$ was also studied using multi-point Pad\'{e} approximants in Ref.~\cite{Pasztor:2020dur}.

On the other hand,
observables related to deconfinement have been studied
much less at non-zero chemical potential. 
The first study to compute the leading Taylor coefficient
of the static quark free energy in $\mu_B^2$ was
Ref.~\cite{DElia:2019iis}, where it was noted that
this coefficient develops a peak near the chiral crossover 
temperature.
In Ref.~\cite{Borsanyi:2024dro} the chiral crossover
temperature was compared to the deconfinement temperature,
defined as the peak position of the static quark entropy.
The volume dependence of both aspects were studied
including their leading $\mu_B^2$ dependence.
It was found that chiral and deconfinement related observables have  
different behaviors. First, deconfinement properties have a milder volume 
dependence. Second, the curvature of the crossover line is slightly larger
for the deconfinement transition. Finally, the deconfinement transition 
appears to get broader with increasing $\mu_B$ (at least up to leading 
order in $\mu_B$), while the width as well as the strength of the chiral transition are approximately constant. 

The main goal of this
work is to further study the deconfinement properties 
of the QCD medium at non-zero $\mu_B$, by extending the
accessible range  in baryochemical potential. 
By calculating the 
Taylor coefficients of the static quark free energy to
order $\mu_B^8$ and expanding around zero, we cover a range up to 
$\mu_B \approx 400$~MeV.

We simulate lattices with a fixed number of 
time-slices, $N_t=8$. However, we expect cut-off
effects to be very small. This expectation is based
on our previous work with the same discretization~\cite{Borsanyi:2023wno}, where
we showed that the value of the fluctuations of the baryon number (up to eighth order) for this discretization are very close to the continuum limit.

In order to 
achieve a good signal for the $\mu_B^8$ coefficient, 
we carry out our study in a smaller 3-volume $LT=2$. 
We showed in Ref.~\cite{Borsanyi:2024dro} 
that quantities related to the Polyakov loop have 
much milder finite volume effects than chiral observables. 
The crossover temperature in the volume we simulate is about $10$~MeV 
higher than its infinite volume value. 
Compared to the infinite volume chiral transition temperature, it is only $5$~MeV higher.
We thus expect the
finite volume effects in our study to be small, though not
completely negligible.

An important technical aspect of drawing phase diagrams in the $T-\mu_B$ plane is how one handles strangeness. In this paper, we study both the phenomenologically more relevant case of zero strangeness expectation value ($n_S=0$, strangeness neutrality) as well as the technically easier case of zero strangeness chemical potential ($\mu_S=0$).

Our final results are the deconfinement transition temperature and the width of the deconfinement transition as a function of the chemical potential.
For the strangeness neutral case they are shown in Fig.~\ref{fig:phase_diagram}. On the left, the deconfinement crossover temperature calculated in this work is compared to the chiral crossover temperature from Ref.~\cite{Borsanyi:2020fev} and an estimate of the
freeze-out parameters in heavy ion collisions by the
STAR collaboration~\cite{STAR:2017sal} and the parametrization 
in Ref.~\cite{Andronic:2017pug}. 
On the right
we show the chiral transition width from Ref.~\cite{Borsanyi:2020fev} and the deconfinement width from this work. Already at $\mu_B=0$ the deconfinement transition is much broader. Even more interestingly, as a function of the baryochemical potential, the deconfinement transition broadens, while the width of the chiral transition stays roughly the same. This contrasting behaviour was already observed in our earlier work~\cite{Borsanyi:2024dro},
but only to leading order in $\mu_B$. Here, we observe
the same behavior in a large range of chemical potential. 
Our results disfavor the existence of a deconfinement critical endpoint below $\mu_B \approx 400$MeV.
As will be shown in Section ~\ref{sec:str_neutr}, the same 
statement is also true for zero strangeness 
chemical potential. However, in that setup, we see
the reverse trend of the width. Above $\mu_B \approx 300$MeV the
width of the deconfinement transition gets smaller again. This is what one expects if a critical endpoint
exists somewhere above $\mu_B = 400$MeV. 

The paper is organised as follows.
In Section \ref{sec:FQ} we motivate our choice of observable to
define the deconfinement temperature.
We work out the details of its extrapolation to finite density through a Taylor expansion in $\mu_B/T$, and describe the renormalization procedure. Cross-checking with ensembles at imaginary $\mu_B$, we
validate the results extrapolated with the Taylor expansion. Note that the extrapolation itself uses solely $\mu_B=0$ ensembles.
In Section \ref{sec:line} we use the extrapolated static quark free energy to
determine the deconfinement temperature up to $\mu_B=400$~MeV
and characterize the strength of the transition. We also compare the cases with and without imposing strangeness neutrality. Finally, we summarize our conclusions 
in Section \ref{sec:conclusions}.

\section{\label{sec:FQ}Static quark free energy at finite density}

The main observable calculated in this work is the
static quark free energy $F_Q$, that is based on the
expectation value $P$ of the Polyakov loop:
\begin{equation}
P = \frac{1}{3}\frac{1}{V} \left\langle \sum_{\vec x\in V} 
\mathrm{Tr}\,\prod_{t=0}^{N_t-1} U_0(t,\vec x)
\right\rangle.
\label{eq:ploop}
\end{equation}
Here the product refers to the multiplication of the 
temporal link variables ($U_t$) of the gauge field along
the entire temporal direction of the lattice that consists of $N_t$ time slices.
Thus, a Wilson line is formed and its normalized trace is the Polyakov loop.
It is averaged over the three-volume $V$ as well as over the ensemble
of gauge configurations.

From $P$, the static quark free energy is then defined as
\begin{equation}
    F_Q = -T \log P.
\label{eq:FQ}
\end{equation}

The Polyakov loop has been a popular observable to study
deconfinement for a long time. This is especially true in the quenched
approximation where the deconfinement
transition is characterized by the spontaneous symmetry breaking of the
center symmetry of the gauge group. The Polyakov loop acts as an order
parameter for this symmetry breaking transition, and its susceptibility
$\chi$ diverges with the volume at the transition temperature. 
In full QCD the role of the Polyakov loop is less evident, especially in
the theory with physical masses, where chiral features are dominant.
Yet, it was observed that the Polyakov loop exhibits a significant rise
near the chiral transition temperature. 
With physical quark masses, the deconfinement transition is a very broad crossover and the Polyakov loop susceptibility
does not provide a clean signature for the transition.

In dynamical QCD the main difficulty with the Polyakov loop is its scheme dependence.
It requires a multiplicative renormalization that is fixed through
an ambiguous procedure. Luckily, the scheme dependence reduces
to an additive term in $F_Q(T,\mu_B)$ that depends on the lattice cut-off,
but not on the infrared control parameters, such as $T$ and $\mu_B$.
We work on the details of renormalization in Section \ref{sec:renormalization}.

It follows from the additive nature of the $F_Q$ renormalization that
the static quark entropy
\begin{equation}
    S_Q(T,\mu_B) = - \frac{\partial F_Q(T,\mu_B)}{\partial T} \, ,
    \label{eq:SQ}
\end{equation}
is well defined, and the scheme dependence cancels in the continuum limit.
It was shown that the crossover temperature defined via $S_Q(T)$ approximately coincides with the
chiral transition temperature for various non-zero values of the quark mass, including the physical case \cite{Bazavov:2016uvm}. In the chiral limit $S_Q(T)$ was shown to exhibit a spike at the chiral $T_c$ \cite{Clarke:2020htu}.
In contrast, the susceptibility of
the Polyakov loop could not be used 
to locate the transition because of
its strong scheme dependence.

In a recent work \cite{Borsanyi:2024dro} we determined the peak position of $S_Q$ for several values
of imaginary chemical potential and found that it follows the chiral transition.
For large volumes, we found that the peak in $S_Q(T)$ is at a temperature roughly 5~MeV lower than the disconnected chiral susceptibility, but its
volume dependence was considerably smaller compared to any of the chiral observables. This
was expected, since $S_Q(T)$ probes QCD with static (infinite mass) quarks,
while chiral quantities access the theory at the pion mass scale. By reducing  the volume from the near-infinite choice of $LT=4$, frequently found in the literature, we are able to compute $S_Q$ at higher density by using an eighth order Taylor extrapolation.

In the following we work out the Taylor extrapolation coefficients of the Polyakov loop
and construct the extrapolation of the static quark free energy $F_Q(T,\mu_B)$.
We compute the coefficients both for the expansion with $\mu_S=0$, 
as well as along the strangeness neutral line $n_S=0$. We demonstrate
the convergence of the series by correctly predicting $F_Q$ at imaginary
$\mu_B$ in both cases.
After discussing the renormalization of $F_Q$ we will be in the position to
calculate $S_Q$ at finite density and determine its peak position as
a function of $\mu_B$.

\subsection{Extrapolation of the static quark free energy}

Since the renormalization of $F_Q$ is additive and $\mu_B$ independent, 
the $\mu_B$ derivatives of $F_Q$ do not require further treatment.
We proceed to formulate a generic chain rule to calculate
derivatives with respect to a quark chemical potential $\mu_j$
that corresponds to the up, down or strange quarks, with $j=u,d,s$ respectively.
We will express the observables in temperature units. Thus, our actual 
expansion parameters are $\hat\mu_j=\mu_j/T$.

The chemical potential dependence is encoded in the quark determinants
that appear in the staggered path integral
\begin{equation}
\begin{aligned}
    Z = &\int \mathcal{D} U  e^{-S_g(U)} \cdot
    {\det M_u(U,m_l,\hat\mu_u)}^{1/4} \cdot \\
        & \cdot{\det M_d(U,m_l,\hat\mu_d)}^{1/4} \cdot
    {\det M_s(U,m_s,\hat\mu_s)}^{1/4} 
\end{aligned}
\end{equation}
where $S_g$ is the gauge action, and $m_l = m_u = m_d$, $m_s$
refer to the light and strange quark masses respectively. Up and 
down quarks are assumed to be degenerate.
For a generic treatment we keep the chemical potentials for the
two light quarks separate, in order to allow for non zero isospin or
electric charge chemical potential, though these are not considered in
the present work.

It is useful to write each determinant's expansion as
\begin{eqnarray}
    \log\det M_j^{1/4}(U,m_j,\mu_j) &=&
    \log\det M_j^{1/4}(U,m_j,0) + A_j \mu_j \nonumber\\
    &&
    + \frac{1}{2!}B_j\mu_j^2
    + \frac{1}{3!}C_j\mu_j^3+\dots
    \label{eq:ABCD}
\end{eqnarray}
where there is no summation over the index $j$.

The chain rule that determines the extrapolation of an
arbitrary variable $X$, following Ref.~\cite{Allton:2005gk}, reads: 
\begin{eqnarray}
\partial_j \avr{X}&=&\avr{X A_j}-\avr{X}\avr{A_j}+\avr{\partial_j X}\,,
\label{eq:chain_rule}\\
\partial_k \partial_j \avr{X}&=&
\frac12\avr{X A_j A_k } 
-\frac12\avr{X}\avr{A_j A_k} 
-\avr{X A_j}\avr{A_k} \nonumber\\
&& +\avr{X}\avr{A_k}\avr{A_j} 
+\avr{(\partial_k X) A_j} -\avr{\partial_k X}\avr{A_k}
\nonumber\\
&&
+\frac12\avr{\partial_k\partial_j X}
-\frac12 \avr{X}\avr{\partial_k A_j}\nonumber\\
&&+(j \leftrightarrow k)
\,.
\end{eqnarray}
where $\partial_j$ stands for $\partial/\partial\hat\mu_j$.
Eq.~(\ref{eq:ABCD}) implies that $\partial_k A_j = \delta_{jk} B_j$.
The second equation is obtained through repeated application of the first one.
This chain rule has already been the basis of numerous computations
of various fluctuations of conserved charges in lattice QCD \cite{Allton:2005gk}. 

The application of this procedure to the Polyakov loop is
straightforward, by applying Eq.~(\ref{eq:chain_rule}) separately to 
its real and imaginary parts. This separation is useful when considering their different symmetry 
features. The odd terms $A$, $C$, \dots, as well as $P_I=i\mathrm{Im}~P$
are imaginary and odd under charge conjugation. On the other hand $B$, $D$,
\dots, as well as $P_R=\mathrm{Re}~P$ are even and real.
In C-symmetric simulations (with $\mu_B=0$) where we calculate the Taylor coefficients, all odd expectation values vanish,
e.g. $\avr{A_j}=0$, $\avr{A_j P_R}=0$ and $\avr{B_j P_I}=0$.
Being a purely gauge observable, the Polyakov loop on a given
gauge configuration is independent of the chemical potentials. 
Hence $\partial_j P_R = \partial_j P_I\equiv0$.

A straightforward application of these rules yields up to second order
\begin{eqnarray}
\left.\partial_j\avr{P_R}\right|_{\mu\equiv0} &=&0\,,\\
\left.\partial_j\avr{P_I}\right|_{\mu\equiv0} &=&\avr{A_j P_I}\,,\\
\left.\partial_j\partial_k\avr{P_R}\right|_{\mu\equiv0} &=&
\delta_{jk} \left[ \avr{B_j P_R} - \avr{B_j}\avr{P_R}\right]\nonumber\\
&&\avr{A_j A_k P_R} - \avr{A_j A_k}\avr{P_R}\,,\\
\left.\partial_j\partial_k\avr{P_I}\right|_{\mu\equiv0} &=&0\,.
\end{eqnarray}

These can be combined into the following relation:
\begin{eqnarray}
\frac12 \partial_j\partial_k |\avr{P}|^2 &=&
+\delta_{jk} \left[ \avr{P_R}\avr{B_j P_R} -\avr{P_R}^2\avr{B_j} \right]
\nonumber\\
&&
+ \avr{P_R}\avr{A_j A_k P_R}
- \avr{A_j P_I}\avr{A_j P_I}\nonumber\\
&&
- \avr{A_j A_k} \avr{P_R}^2 \, \, ,
\label{eq:P2}
\end{eqnarray}
which, at the level of the static quark free energy, reads
\begin{eqnarray}
    \partial_j \partial_k F_Q &=& 
-\frac{T}{\avr{P_R}^2}\frac12 \partial_j\partial_k |\avr{P}|^2 \, \, .
\label{eq:FQ2}
\end{eqnarray}
Eqs.~(\ref{eq:P2}) and (\ref{eq:FQ2}) have been first formulated
and used by the Pisa group in Ref.~\cite{DElia:2019iis}. To our knowledge, higher order derivatives have never been addressed. 

For the derivation of higher orders we have set up two independent
automated procedures, so that they can be cross-checked. In the first
approach we have built a computer algebra program (in C language)
that applies the chain rule and the symmetry rules.
In the second setup a Python code has implemented combinatorial 
considerations short-cutting the tedious sequential derivatives.
Both programs generated an identical computer code that was then used to
calculate the numerical values of the coefficients. Examples of the
formulas are given in Appendix~\ref{app:formulas}.

All combinations of quark chemical potentials have been computed.
From these we constructed the generic derivatives (up to eighth order)
in the $\mu_B$-$\mu_S$ basis. Then, we could construct the expansion in
$\mu_B$ both at $\mu_S=0$, as well as on the strangeness neutral line. 
In the latter case, $\mu_S$ itself is subject to a Taylor expansion:
\begin{equation}
\hat\mu_S(\hat\mu_B)=
s_1 \hat\mu_B + s_3 \hat\mu_B^3 + 
s_5 \hat\mu_B^5 + s_7 \hat\mu_B^7 +\dots
\end{equation}
We presented the continuum extrapolations for the first three coefficients
in Ref.~\cite{Borsanyi:2023wno}. The formulas related to this 
expansion have been published by the HotQCD group \cite{Bazavov:2017dus}.

The final expansion of $F_Q$ is given in terms of the $F_n$ coefficients
defined as
\begin{equation}
\begin{aligned}
F_Q(T,\mu_B)&=-\frac{T}{2}\log{|\avr{P}|^2} = \\
&=F_Q(T,\mu_B=0) + T \sum_{n=2,4,\dots} \frac{F_n}{n!} \left(\frac{\mu_B}{T}\right)^n \, \, .
\\
\end{aligned}
\end{equation}

We relate the expansion of $|P|^2$ to that of $F_Q$ with the straightforward
formulas (\ref{eq:FQ_sum})--(\ref{eq:FQ_8})
in Appendix~\ref{app:formulas}.

\begin{figure*}[t] 
\centerline{
\includegraphics[width=0.50\textwidth]{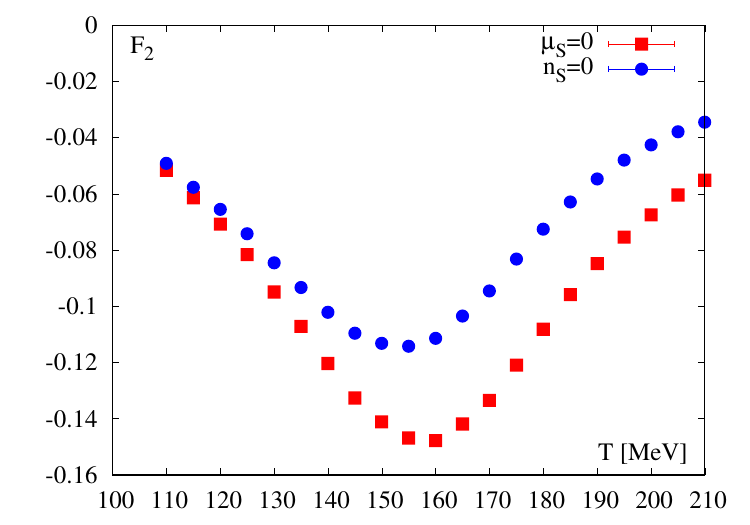}
\includegraphics[width=0.50\textwidth]{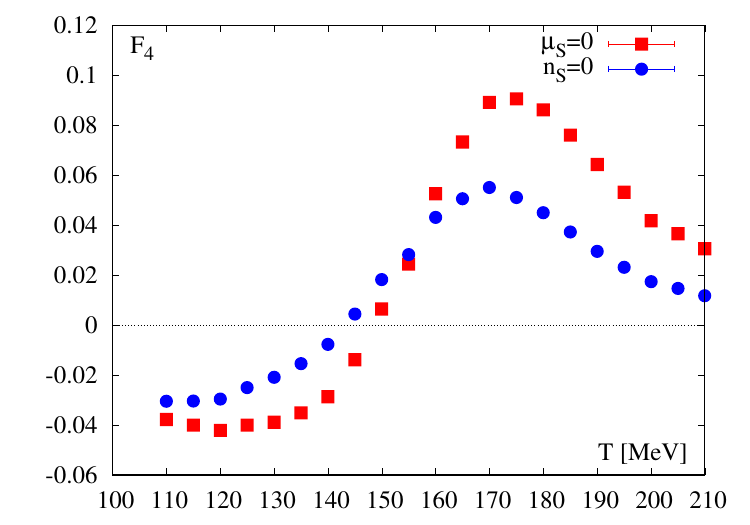}
}
\centerline{
\includegraphics[width=0.50\textwidth]{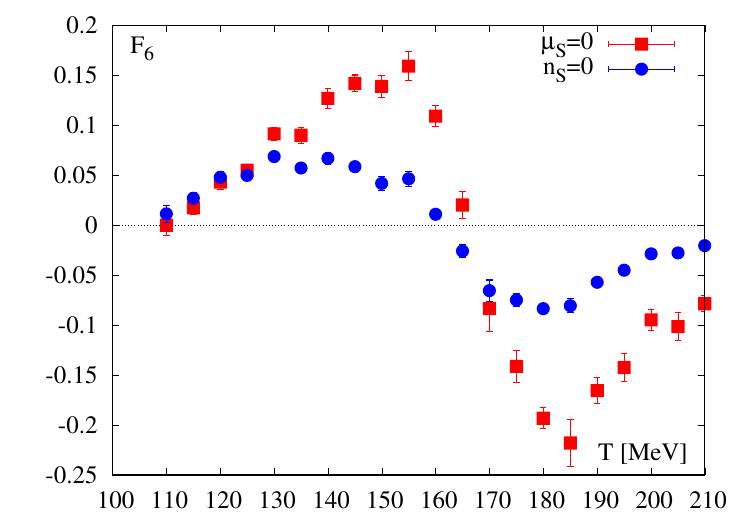}
\includegraphics[width=0.50\textwidth]{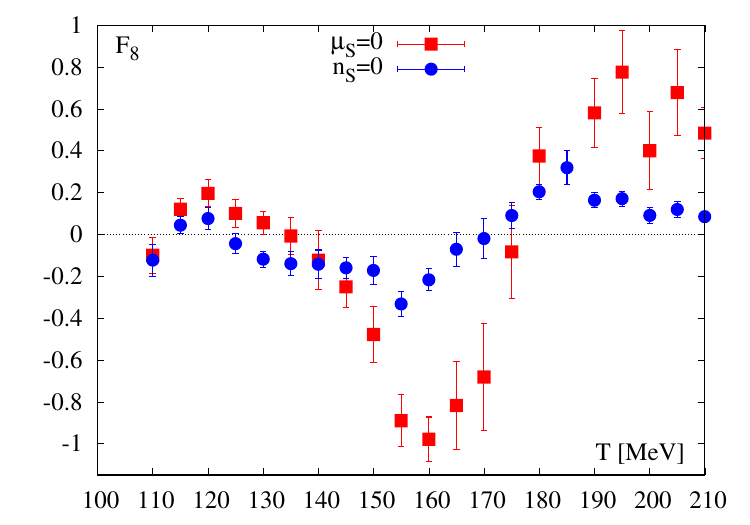}
}
\caption{\label{fig:FQn}
Taylor coefficients of the static quark free energy in two setups: vanishing strangeness chemical potential in red, along strangeness neutral trajectories in blue.
}
\end{figure*}

\subsection{\label{sec:Fnresults}Expansion coefficients from the lattice}

The sign problem appears in high order expansion coefficients in the form
of large cancellations between competing terms. These lead
to signal-to-noise ratios that are suppressed proportionally to the volume, 
by an additional factor for each subsequent order.
This has forced us to reduce the volume to $16^3\times 8$ for
the scope of this study, and to work with extreme statistics of the order of
$\mathcal{O}(10^6)$ configurations.
The accumulated statistics and simulation parameters are listed in 
Appendix~\ref{app:statistics}.
We use ensembles with 4HEX staggered fermions as in
our recent work  \cite{Borsanyi:2023wno}. However, the statistics is increased 
by one order of magnitude.
A further ingredient we employed to enhance the signal-to-noise ratio 
is the application of four HEX smearing steps to the Polyakov loop.
This noise-reducing step is irrelevant in the continuum limit, and does not
alter the sensitivity of the Polyakov loop to the breaking of the
center symmetry. A similar strategy of using smeared
Polyakov loops as noise reduction was used in Ref.~\cite{Borsanyi:2015yka}.

In Refs.~\cite{Borsanyi:2022soo,Borsanyi:2023tdp,Borsanyi:2023wno} 
we have already benefited from the reduced matrix formalism
\cite{Hasenfratz:1991ax} to obtain the 
configuration-specific expansion coefficients $A_j,B_j,\dots$
in Eq.~(\ref{eq:ABCD}) to arbitrary order. The chemical potentials
are consistently defined to correspond to exactly conserved 
charges following the definition in Ref.~\cite{Hasenfratz:1983ba}.

Thus, our approach is unlike the standard strategy to calculate these coefficients in lattice QCD, which relies on Gaussian random vectors \cite{Allton:2002zi,Allton:2005gk}. The reduced matrix formalism
defines a large matrix of size $(6N_s^3)^2$ as means of a
deterministic computation of the fermion determinant. In our case, 
it is a $24576\times 24576$ matrix.
The knowledge of all eigenvalues $\xi_i$ allows the
evaluation of the fermion determinant 
in a closed formula for any quark chemical potential $\mu$
\begin{equation}
\frac{\det M(\mu,m,U)}{\det M(0,m,U)}=e^{-3N_s^3\mu/T} \prod_{i=1}^{6N_s^3} \frac{\xi_i[m,U] - e^{\mu/T}}{\xi_i[m,U] -1}\,.
\end{equation}
At vanishing $\mu$ one can readily obtain the coefficients by following
simple derivation rules. For low orders one obtains: 
\begin{eqnarray}
    A&=&\frac{i}{4} \mathrm{Im}~\sum_i \frac{1}{1-\xi_i} \, \, ,\\
    B&=&\frac{1}{4} \mathrm{Re}~\sum_i \frac{-\xi_i}{(1-\xi_i)^2} \, \, ,\\
    C&=&\frac{i}{4} \mathrm{Im}~\sum_i \frac{\xi_i(1+\xi_i)}{(1-\xi_i)^3} \, \, ,\\
    D&=&\frac{1}{4} \mathrm{Re}~\sum_i \frac{\xi_i(\xi_i^2+4\xi_i+1)}{(1-\xi_i)^4}  \, \, .
\end{eqnarray}

We have one such matrix for each of the two masses
for every configuration. That means
42 million separate diagonalizations for this paper altogether.
We compute the eigenvalues using a dense
linear algebra package, MAGMA \cite{tdb10,MR1484478,BrownATD20}
on the LUMI supercomputer in Finland.

We present the $F_n$ coefficients computed on our $16^3\times 8$ lattice in Fig.~\ref{fig:FQn}.
The red symbols correspond to the expansion in $\mu_B$ with $\mu_S=0$,
the blue data refer to the strangeness neutral case $(n_S=0)$.
The second order coefficient was previously calculated in
a combination of imaginary-$\mu_B$ and Taylor approach with $\mu_S=0$
in Ref.~\cite{DElia:2019iis} (using the opposite sign convention).
The results are very similar, even though the volume and the action are different. The other seven coefficients are
presented here for the first time.

\subsection{\label{sec:crosscheck}Consistency with imaginary $\mu_B$ data}

Before we move on to the applications we perform a cross-check
of the coefficients by comparing them to imaginary $\mu_B$ data.
We obtained the $F_n$ coefficients entirely from $\mu_B=0$ simulations.

\begin{figure*}[t]
\centerline{
\includegraphics[width=0.50\textwidth]{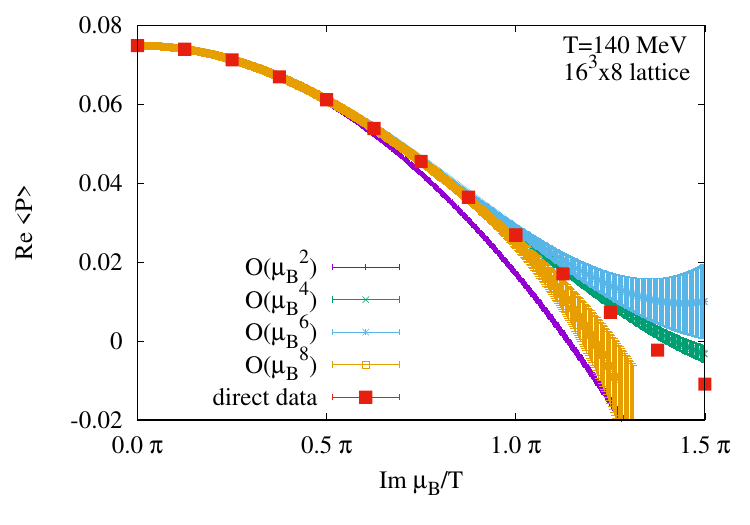}
\includegraphics[width=0.50\textwidth]{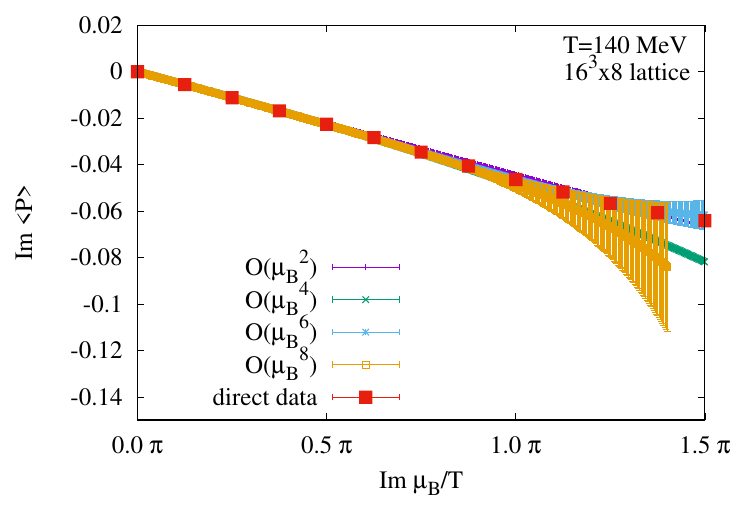}
}
\caption{\label{fig:Pextra}
The Taylor coefficients are used to extrapolate
the bare Polyakov loop expectation value
from $\mu_B=0$ to imaginary values of the baryochemical potential.
(Left: real part, Right: imaginary part.)
We demonstrate the convergence of the series at $T=140$~MeV by
comparing orders of the extrapolations (bands) to 
directly measured expectation values at $\mathrm{Im}~\mu_B>0$ from a separate set of simulations.
}
\end{figure*}

For the sake of this cross-check we simulated at eight imaginary
values of $\mu_B$ at $T=140$~MeV, measuring the Polyakov loop in each run.
By substituting imaginary values into our Taylor expansion we could make
predictions on the imaginary $\mu_B$ runs' outcome.

In Fig.~\ref{fig:Pextra} we compare these predictions to the direct
imaginary-$\mu_B$ simulations. The expansion produces reasonable predictions already at leading order as far as $\mu_B/T \sim i \pi/2$, and with 
higher orders it is compatible with the direct result up to $\mu_B/T \sim i \pi$.
In that interval the eighth order gives only a small correction, thus
demonstrating convergence.

\begin{figure}[t]
\centerline{
\includegraphics[width=0.5\textwidth]{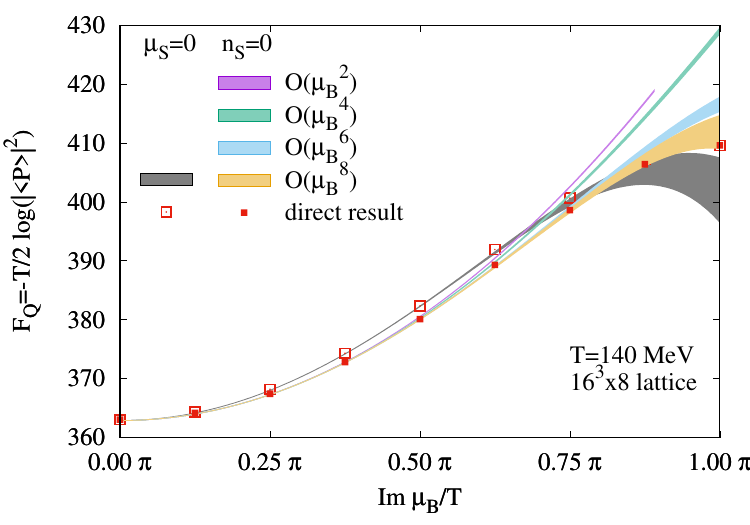}
}
\caption{\label{fig:FQN_140}
The Taylor coefficients are used to extrapolate
the bare static quark free energy ($F_Q)$ 
from $\mu_B=0$ to imaginary values of the baryochemical potential.
We show the comparison both for the $\mu_S=0$ as well as the $n_S=0$
case. The respective expansion reproduces the direct data very well.
}
\end{figure}

As an additional check we compute the bare $F_Q(\mu_B)$ for imaginary
chemical potentials. Since renormalization will only apply a constant shift
to the entire plot, it plays no role for this comparison.
In order to obtain direct data along the strangeness neutral
line at imaginary $\mu_B$, we reweighted each ${\rm Im}~\mu_B>0$ ensemble to the specific
imaginary $\mu_S$ where $n_S=0$. The target $\mu_S(\mu_B)$ was also computed
using reweighting, with reweighting factors spread between 0.5 and 2.
Even in the $n_S=0$ scheme, our error bars are smaller than the
symbol size. We stress that reweighting was only applied to the imaginary 
$\mu_B$ data and only for the sake of this crosscheck. We will refer
to the resulting $F_Q$ for both schemes as direct data, as opposed to
the Taylor expansion, that is based on the $\mu_B=0$ ensemble. We show the comparison in Fig.~\ref{fig:FQN_140}. 
For the phenomenologically
more relevant strangeness neutral case ($n_S=0$) we show four subsequent
orders, while for the $\mu_S=0$ expansion we plot the highest order only.
Although the two expansion schemes clearly differ, each set of coefficients
reproduce the respective direct data set. Interestingly, the strangeness
neutral extrapolation reaches a better agreement with direct data, 
and has smaller statistical noise. We observe a monotonic convergence
for $\mu_B/T$ up to $i\pi$.

\subsection{Renormalization at $\mu_B=0$\label{sec:renormalization}}

The renormalization of the Polyakov loop, or that of $F_Q$ can be
discussed independently of our extrapolation, and its details 
have already been worked out in the literature. The various 
methods can be classified into three approaches:
\begin{enumerate}[i)]
\item The direct method exploits the fact that the difference of
$F_Q$ between two temperatures is finite, if the two simulations
used the same bare parameters \cite{Gupta:2007ax,Borsanyi:2015yka}.
\item The static quark potential ($V_{Q\bar Q}$) uses $T=0$ simulations
and relates the renormalization of $V_{Q\bar Q}$ to that of $F_Q$
\cite{Gupta:2007ax,Cheng:2007jq}
or to the finite temperature heavy quark potential at small distances \cite{Kaczmarek:2002mc}.
\item The Polyakov loop is finite at fixed finite gradient flow time
\cite{Petreczky:2015yta,Datta:2015bzm,Bazavov:2016uvm}.
\end{enumerate}

\begin{figure}[t]
\centerline{
\includegraphics[width=0.5\textwidth]{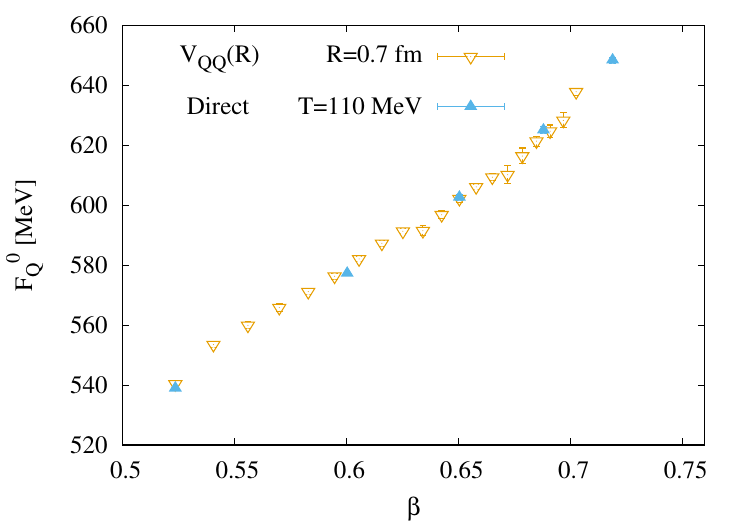}
}
\caption{\label{fig:renormalization}
We use two different approaches to the renormalization of the
Polyakov loop. The full triangles correspond to the direct method i),
the open triangles is based on zero temperature simulations of
Wilson loops ii). The two data sets are shifted by a constant value
so that their agreement is apparent.
}
\end{figure}

In Fig.~\ref{fig:renormalization} we show the gauge coupling ($\beta$)
dependence of the renormalization constant $F_Q^{(0)}(\beta)$
for schemes i) and ii).
For option iii) the flow time should be at a shorter length scale than
other physical scales, yet it should also be above the scale of the
lattice spacing. Given that the separation of these scales for $N_t=8$ is not
clear enough to endorse a suitable choice for flow time, we did not consider this approach.

For i) we simulated a fixed temperature $T^*=110$~MeV on a range
of lattice resolutions $N_t=8,10,12,14$ and 16. The lattice spacing
of the $N_t=16$ ensemble at $T^*$ corresponds to $220$~MeV for $N_t=8$.
Thus, these five
ensembles conveniently cover the entire range where renormalization
is required. Then, $F_Q^{0}(\beta)$ is obtained as (an arbitrary constant plus)
the measured bare $F_Q$ from this setup.
Other choices for $T^*$ are possible, as the resulting renormalized
$F_Q$ would then differ by $\mathcal{O}(a^2)$ 
discretization errors only. However, larger values of $T^*$ would force us
to use $N_t=6$ lattices that may have insufficient resolution. At smaller $T^*$ the Polyakov loop is even smaller since then we are deep in the
confined phase. It is a challenge then to keep the signal-to-noise ratio
high enough, let alone that in that case an $N_t=20$ ensemble would
also be required to span the full $\beta$ range.

For ii) we simulated one $32^3\times64$ dedicated ensemble for
each finite temperature run with a statistics of $\approx 1000$ configurations in each. The 4HEX-smeared Wilson loops were calculated
with additional spatial smearing (128 stout steps). We followed
the standard procedure to extract and fit the effective mass
from the time separation dependence of the Wilson loop and interpret
these as $V_{Q\bar Q}(R)$, where $R$ is the fixed spatial separation.
Due to the 4HEX smearing the Cornell potential is distorted at
small distance, so we fitted a generic rational function on $V_{Q\bar Q}(R)$ to interpolate at each gauge coupling
to the same physical scale.
Using a fixed $R=0.7$~fm scale $V_{Q\bar Q}(R)/2$ gives the
$\beta$-dependent renormalization factors. The factor 1/2 comes
from the forward and backward occurrence of the time-like line
within the Wilson loop.

The $F_Q^0(\beta)$ counterterms slightly differ in the two methods.
Their deviation is considered as one of the sources of systematic errors on $S_Q(T,\mu_B)$.

\section{\label{sec:line}The phase diagram}

\subsection{Deconfinement at finite $\mu_B$}

In the previous section we have worked out the expansion coefficients
of the static quark free energy $F_Q$ in $\mu_B/T$ and also computed
it at zero chemical potential with two different renormalization methods.
In short, we express $F_Q(T,\mu_B)$ as the difference of two terms
\begin{eqnarray}
F_Q(T,\mu_B) &=& F_Q^{\rm bare} (T,\mu_B)  - F_Q^{c.t.} (T).\label{eq:FQsub}\\
F_Q^{\rm bare}(T,\mu_B) &=& F_Q^{\rm bare} (T,0) + T \mkern-8mu\sum_{n=2,4,\dots} \frac{F_n}{n!} \left(\frac{\mu_B}{T}\right)^n
\label{eq:FQbare}
\end{eqnarray}
Here the counterterm $F_Q^{c.t.}$ that was originally
defined as a function of the gauge coupling ($\beta$) is rewritten
as a function of temperature ($T=1/(a N_t)$ using the $a(\beta)$
scale setting. The $a(\beta)$ function
for this action was introduced in Ref.~\cite{Borsanyi:2023wno}.

\begin{figure}[t]
\includegraphics[width=0.48\textwidth]{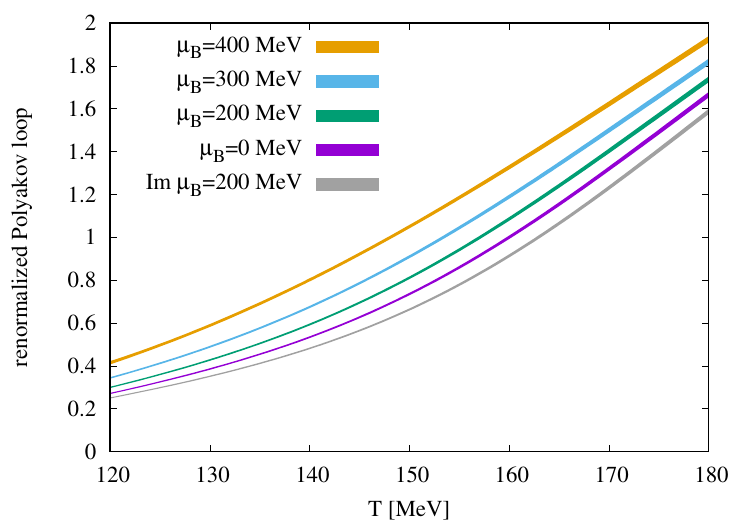}
\includegraphics[width=0.48\textwidth]{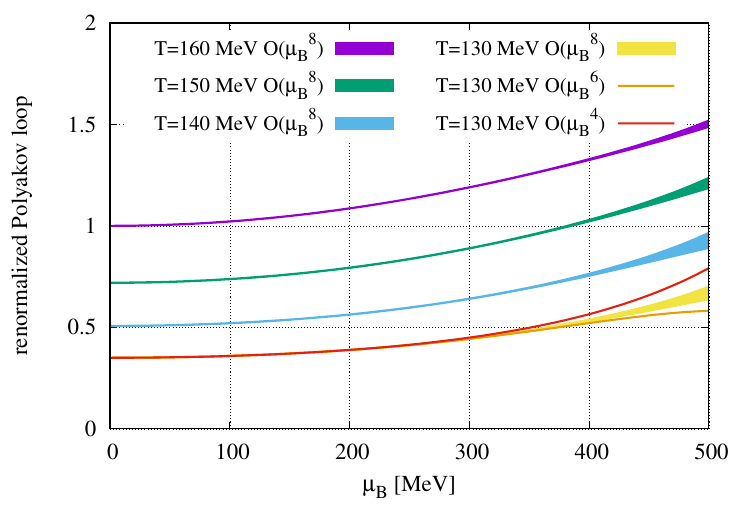}
\caption{\label{fig:ploop_series}
Top:
Renormalized Polyakov loop ($P^r$) as a function of the temperature for various real chemical potentials, in the scheme defined
by $P^r(T=160~\mathrm{MeV},\mu_B=0)=1$. The extrapolations
are to eighth order in $F_Q$. We added the result with
an imaginary chemical potential for comparison (gray line).
Bottom: The same information, shown as a function
of $\mu_B$ with fixed $T$. For one of the temperatures we show
the fourth and sixth order results. The sixth order does not visibly differ
from the full computation below $\mu_B=350$~MeV.
}
\end{figure}

We start with the presentation of the renormalized Polyakov loop, 
since most readers are more familiar with this observable than
$F_Q$ or $S_Q$. 
Since we have already established $F_Q$
at finite $\mu_B$ we define
the renomalized Polyakov loop as 
$P^r=\exp{-(F_Q-\Delta)/T}$,
where $\Delta$ sets the scheme. This means that for observables
in this paper we always expand $F_Q$, not $P_R$ or $P_I$.
We choose $\Delta$ such that $P^r(T=160~\mathrm{MeV},\mu_B=0)=1$
as we did in Ref.~\cite{Borsanyi:2024dro}. The Polyakov loop 
curves are shown in Fig.~\ref{fig:ploop_series}.
In the bottom panel of this figure, we also show the fourth, sixth and eighth order curves for $T=130$~MeV. The expansion is under control below $\mu_B \approx 400$~MeV.

\begin{figure}[t]
\includegraphics[width=0.5\textwidth]{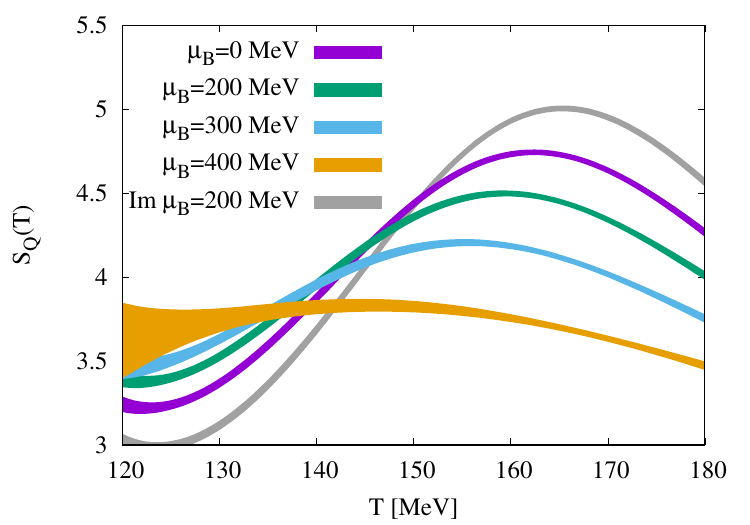}
\caption{\label{fig:SQ}
The static quark entropy ($S_Q(T)$) extrapolated
to various chemical potentials. The errors in this plot are statistical
only. We added one imaginary chemical potential (gray band) to better visualize the trend in $\mu_B$.
}
\end{figure}

One could be tempted to define a proxy for the deconfinement
transition via a constant value of $P^r$ or $F_Q$.
While one can construct such contours with little effort, their
curvature will be less than that of the chiral transition line.
We have shown this in Ref.~\cite{Borsanyi:2024dro}, where $F_Q$
and the chiral observables were compared at imaginary $\mu_B$. Along the transition line $T_c(\mu_B)$, $F_Q$ showed a statistically significant slope as a function of $\mu_B^2$.

In this work, however, we focus on $S_Q(T,\mu_B)$. Its maximum
corresponds to the inflection point of $F_Q(T)$.
It is challenging to reach the precision needed to extract
the second derivative with numerical differentiation. In
Ref.~\cite{Borsanyi:2024dro} we have already accomplished this for several
volumes and chemical potentials. We modelled the
temperature dependence with a rational function, from which the inflection point was determined.

In Fig.~\ref{fig:SQ} we show $S_Q(T)$ as defined in Eq.~(\ref{eq:SQ}).
The derivative was calculated by choosing
a suitable interpolation of the two terms in Eq.~(\ref{eq:FQsub}), keeping $\mu_B$ fixed in MeV.
We also show $S_Q(T)$ for one imaginary value of $\mu_B$ for comparison.
The $S_Q(T)$ curves show two trends very clearly. First, the
change in the peak position indicates the chemical potential dependence
of the transition temperature. 
Second, the width of the peak increases as $\mu_B$ grows, indicating a weakening of the deconfinement transition. At the largest value of $\mu_B$, the curve is the flattest. 
This feature contributes to the difficulty in determining the deconfinement temperature at large 
$\mu_B$, in addition to the growing errors
from the extrapolation.
On the other hand, in the opposite
(imaginary) direction, the peak is higher and
narrower, eventually becoming
singular in the Roberge-Weiss point \cite{Bonati:2016pwz}.

\subsection{Systematic analysis}

In the following we describe the analysis procedure that allows
us to draw the transition line in the phase diagram based on $S_Q$.

Renormalization introduces a lattice-spacing dependent
shift of $F_Q$. This function is interpolated to the actual spacing
that realizes a given temperature ($a=1/(N_t T)$ with $N_t=8$).
The interpolation of the same data to various temperature points
introduces correlations between temperatures. To avoid this, we
model the renormalization function $F_Q^{c.t.}(T)$ 
in Fig.~\ref{fig:renormalization} with a polynomial, independently from $F_Q^{\rm bare}(T)$.

$F_Q^{\rm bare}$ is calculated at each value of $\mu_B$ (not $\mu_B/T$) as a function of the temperature,
by simply summing the Taylor series to eighth order (Eq.~\ref{eq:FQbare}). This extrapolated
data correspond to the would-be simulation result as if we simulated at finite $\mu_B$. Each
data point corresponds to a definite pair of $T$ and $\mu_B$. 
There is no statistical correlation between results at different temperatures,
since each was obtained from a single $\mu_B=0$ ensemble. We fit $F_Q^{\rm bare}$ 
with a rational function for every $\mu_B$ separately, in order to calculate its $T$-derivative, namely $S_Q(T)$. 
In this work we used (3,2), (i.e., cubic over quadratic)
or (3,3) (i.e., cubic over cubic) rational functions.
The renormalized $S_Q(T)$ function is simply the difference of the
temperature derivative of the rational function and that of $F_{Q}^{c.t.}(T)$.
The maximum of $S_Q(T)$ for fixed $\mu_B$ is found without further modelling of the function.

The determination of the inflection point and the renormalization of $F_Q$ are
both procedures where ambiguous choices have to be made. We identified five such
steps, and we use two alternative versions for each. Thus, there are a total of
32 $S_Q(T,\mu_B)$ curves for each $\mu_B$ value. In short, they include: i)
whether to apply a polynomial model for $F_{Q}^{c.t.}(\beta)$ or for
$F_{Q}^{c.t.}(T)$ using the scale function $a(\beta)$ to convert between the two
options; ii) whether to fit these counterterm functions with third or fourth
order polynomials; iii) whether to use the zero or finite temperature data set
to renormalize; iv) two different ranges for the rational model of the bare
$F_Q$; and finally v) whether to use (3,2) or (3,3) rational approximations in
$T$ for $F_Q^{\rm bare}(T,\mu_B)$. The systematic error is calculated with  the
histogram method of Ref.~\cite{Borsanyi:2020mff}.

Once $S_Q(T,\mu_B)$ is known, its peak position gives the deconfinement
crossover curve in the phase diagram. This is shown in the left panel of
Fig.~\ref{fig:phase_diagram}, with all of the systematic errors included.
Furthermore, we can define the width of the transition as
\begin{equation}
\Delta T(\mu_B) = \left.\sqrt{
- \left. S_Q(T,\mu_B)\,\middle/\,\frac{\partial^2 S_Q(T,\mu_B)}{\partial T^2}\right.}\right|_{T=T_c(\mu_B)}\hspace*{-15mm}~
\label{eq:width}
\end{equation}
In Ref.~\cite{Borsanyi:2020fev} an analogous definition was given for the full chiral susceptibility.
While in Ref.~\cite{Borsanyi:2020fev} we constructed a proxy using the chiral condensate, here
data quality allows us to directly compute the second derivative of $S_Q$ at $T_c(\mu_B)$.
For simplicity, we re-fitted the $S_Q(T,\mu_B)$ function at the peak of $S_Q(T)$ in two different
ranges ($T_c\pm10$~MeV and $T_c\pm15$~MeV). With this sixth step the number of analyses totals 64.
The resulting curve in the strangeness neutral case is shown in the right panel of Fig.~\ref{fig:phase_diagram}. Already at $\mu_B=0$, the width defined with $S_Q$ is considerably
larger than the chiral width.
Furthermore, unlike the chiral width, which is approximately 
a constant at small chemical potentials, the deconfinement width gets larger with increasing $\mu_B$, indicating a weakening crossover transition. One possible interpretation
of this fact is that it is due to the increased distance from
the Roberge-Weiss critical point~\cite{Bonati:2016pwz}.
A widening crossover implies that the existence of a 
critical point in this region is disfavored by the 
results of our analysis.

\subsection{The role of strangeness neutrality}\label{sec:str_neutr}

\begin{figure}[t]
\includegraphics[width=0.48\textwidth]{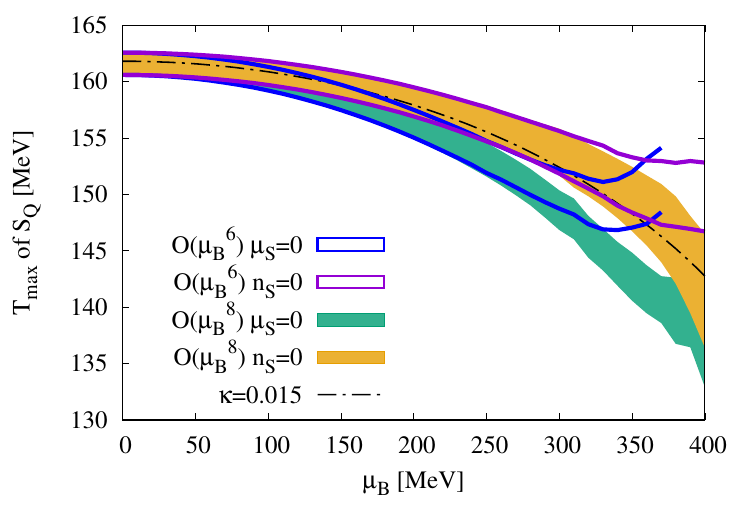}
\includegraphics[width=0.48\textwidth]{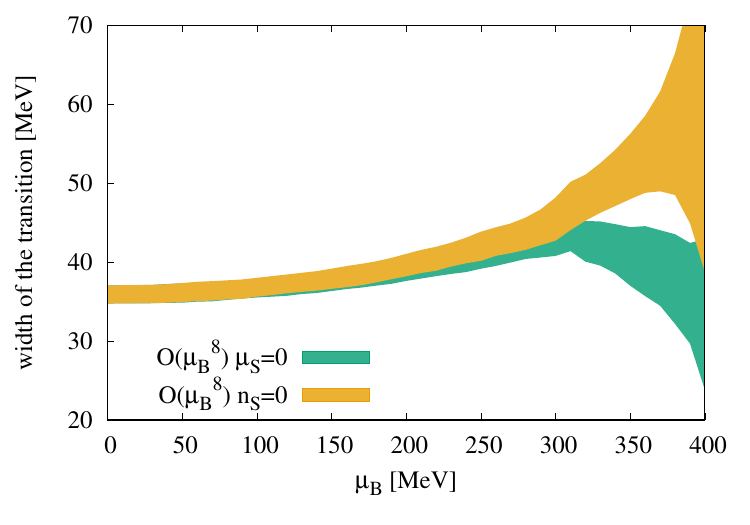}
\caption{\label{fig:neutralvsnot}
Top:
The maximum position of $S_Q(T)$ for fixed baryo-chemical potential ($\mu_B$). We show both setups with ($n_S=0$) and without ($\mu_S=0$) strangeness neutrality. We quote a rounded value of the curvature
of the chiral transition and show the corresponding transition line for further guidance.
We also show the sixth order result on the phase diagram. The eighth order is clearly significant above 350~MeV.\\
Bottom:
The width of the transition defined as Eq.~(\ref{eq:width}). While the width is similar for the strangeness neutral and in the $\mu_S=0$ case we see a hint for a strengthening for $\mu_S=0$ at the highest chemical potentials.
 }
\end{figure}

We finally discuss the impact of strangeness neutrality. In Fig.~\ref{fig:neutralvsnot} we show
the deconfinement transition line with the two standard choices: the simple extrapolation in $\mu_B$ keeping $\mu_S=0$,
and the strangeness neutral case $n_S=0$. The latter was shown in our final result in Fig.~\ref{fig:phase_diagram}. 
Since the simpler condition $\mu_S=0$ is often used in lattice studies, 
we present a comparison here. (For recent estimates of 
the critical endpoint's possible location in the 
$\mu_S=0$ scheme see Refs.~\cite{Clarke:2024ugt, Basar:2023nkp, Hippert:2023bel}).

Similarly to the chiral transition line, the maximum of $S_Q$ is also well
approximated by the quadratic ansatz
\begin{equation}
    T_c(\mu_B) = T_c(0) \left( 1 - \kappa\, \frac{\mu_B^2}{T_c^2(\mu_B)}\right)
\end{equation}
that we plot together with our lattice result to guide the eye. We used the rounded value of $\kappa=0.015$.
For specific extrapolation schemes, more precise values have been determined from lattice simulations
\cite{Cea:2015cya,Bonati:2015bha,Bellwied:2015rza,Bazavov:2018mes,Borsanyi:2020fev,Ding:2024sux}.
The transition lines in Fig.~\ref{fig:neutralvsnot} indicate a stronger curvature for the $\mu_S=0$ case,
as expected.

The second panel in Fig.~\ref{fig:neutralvsnot} shows the differences in the width of
the $S_Q$ peak. While we see a steady increase in the width, i.e. a weakening
of the deconfinement transition in the $n_S=0$ case, the $\mu_S=0$ shows a remarkable
turn of this trend above 300 MeV. Given this behaviour in Fig.~\ref{fig:neutralvsnot}
one can speculate if the width continues to drop, perhaps even down to zero.
This would be consistent with the critical end point existent scenario.
One could speculate that such reversal of the trend can also happen to the $n_S=0$ width at a higher $\mu_B$, but the error on our results at 
$\mu_B>400~\mathrm{MeV}$ are two large to explore this 
possibility.

Note that this effect of the transition first 
weakening and then strengthening again is reminiscent
of result from early work on the phase diagram
on coarse staggered lattices~\cite{Fodor:2004nz, Giordano:2020uvk, Giordano:2020roi}.

Finally, we address the robustness of the present computation.
We use the eighth order of the Taylor series, which may seem a very arbitrary decision.
In fact, the order of the expansion was only limited by the statistics.
In Fig.~\ref{fig:neutralvsnot} we added the $\mathcal{O}(\mu_B^6)$ order version
of our extrapolated transition lines, too. Up to 300~MeV the effect of the eighth order
is invisible, and only gets statistically significant above $\mu_B\simeq350$~MeV.
For the $\mu_S=0$ scheme the onset of the eighth order appears 50~MeV earlier.
It is safe to claim that up to $\mu_B\le 300~\mathrm{MeV}$ the systematic errors
from the truncated extrapolation can be completely neglected, and we give a fair description
of finite volume QCD at finite $\mu_B$. At the same time, the 
eighth order is the lowest possible truncation in the range 300-400~MeV. Lower orders cannot give reliable results.
The difference in the width between the $n_S=0$ and 
$\mu_S=0$ schemes is an effect that is visible only if the eighth order is included.

\section{\label{sec:conclusions}Conclusions}

We studied the crossover line on the QCD phase diagram
in an unprecedented range of chemical potentials.
We complemented previous lattice results on the chiral aspects of the QCD transition
by focusing on observables that are associated with deconfinement.

We calculated the Polyakov loop and the static quark 
free energy in hot and dense QCD as functions of $T$ 
and $\mu_B$. Direct simulations at finite $\mu_B$ are hindered
by the sign problem, but various expansion techniques exist
to extrapolate the results to finite density. Here we employed
the Taylor method using an eighth order expansion in the 
baryochemical potential-to-temperature ratio $\mu_B/T$. 

The Taylor method, which is often used to extrapolate the QCD pressure,
is known to suffer from a remnant sign problem: the positive and negative
terms (e.g. those we list in Appendix \ref{app:formulas}) exhibit a cancellation.
This sign problem worsens with each subsequent order of the expansion with a volume
dependent factor. Although the sixth and eighth order coefficients
of the pressure are often studied in the literature, the errors are large.

The novelty in this work is the extension of the Taylor scheme to the
Polyakov loop and related observables. While this step posed no conceptual
challenge, the derivation and implementation of the high order coefficients
would have been a formidable task without the use of a computer algebra system that
we developed for the purpose. We note that no new lattice measurements had
to be implemented in the simulation code other than
the Polyakov loop itself and the standard routines that are required for the expansion of the pressure.

Yet, the most important ingredient that has allowed us to reach 
400~MeV in the baryo-chemical potential was the selection of a simulation
volume that keeps the sign problem under control. 
Indeed, the errors on the expansion coefficients in this work are a fraction of those
that are often used in large volume simulations (e.g. for the pressure).
Finite volume effects are not negligible, but were quantified in a previous work, and found to be small.
While chiral observables suffer from large finite volume corrections,
the Polyakov loop is much less sensitive. 
The aspect ratio of $N_x/N_t=L\cdot T=2$ is a choice that allows
a wider opening of our window in the QCD phase diagram.

Simulating with extreme statistics ($>$ 1 million configurations)
at each temperature in a finite volume we calculated the bare
Polyakov loop in the entire range covered by the Beam Energy Scan
program in collider mode. This bare Polyakov loop would have been
the direct result if simulations at finite $\mu_B$ were possible.

We defined the transition line as the peak of the static quark entropy.
For this we used a renormalization procedure that
has already been extensively discussed in the literature.

The transition line in this work could be extended up to 400~MeV in
the baryo-chemical potential. It closely follows the chiral crossover line,
though the latter is known with larger errors and in a smaller range.

With this result the phenomenological freeze-out line can be compared to
the deconfinement line in a longer range than before. In particular, we cover the full range
of the STAR freeze-out determination of Ref.~\cite{STAR:2017sal}.

The curvature of the static quark entropy at its peak can be used to define a 
width parameter for the deconfinement crossover. We showed that up to $\mu_B \approx 400$~MeV, 
the width parameter is getting larger at larger $\mu_B$. This disfavors the existence of a deconfinement 
critical endpoint in this regime. 

Finally, we compared the phenomenologically more relevant case of strangeness neutral matter with the theoretically more easily tractable case of zero strangeness chemical potential. While the curvature of the $T_c(\mu_B)$ curve is
slightly smaller for the strangeness neutral case, the width parameters are
consistent (within error) up to $\mu_B \approx 300$~MeV. In the range $300~\rm{MeV} < \mu_B < 400$~MeV, there is a difference in the width parameters, with
the $\mu_S=0$ curve apparently turning around, and showing a narrowing of the
transition (at least to this order in the Taylor expansion). This behavior might
be caused by a critical endpoint beyond $\mu_B=400$~MeV. One might speculate
that such a narrowing can also happen to the $n_S=0$ transition, but at higher
values of $\mu_B$. Unambiguously deciding whether this is the case calls for
further investigations.

\begin{acknowledgments}
The project was supported by the BMBF Grant
No. 05P21PXFCA. This work is also supported by the
MKW NRW under the funding code NW21-024-A. Further
funding was received from the DFG under the Project
No. 496127839. This work was also supported by the
Hungarian National Research, Development and Innovation
Office, NKFIH Grant No. KKP126769.
This work was also supported by the NKFIH excellence
grant TKP2021{\textunderscore}NKTA{\textunderscore}64.
This work is also supported by the Hungarian National Research,
Development and Innovation
Office under Project No. FK 147164.
The authors gratefully acknowledge the Gauss Centre for
Supercomputing e.V. (\url{www.gauss-centre.eu}) for funding
this project by providing computing time on the GCS
Supercomputer Juwels-Booster at Juelich Supercomputer
Centre. 
We acknowledge the EuroHPC Joint Undertaking for awarding this project access to the EuroHPC supercomputer LUMI, hosted by CSC (Finland) and the LUMI consortium through a EuroHPC Extreme Access call.
\end{acknowledgments}

\appendix
\section{\label{app:statistics}Statistics and parameters}

\begin{table}[h!]
\begin{tabular}{|c|c|c|c|r|}
\hline$T$ [MeV] & $\beta$ & $m_l$ & $m_s$ & \# configs\\
\hline
110 & ~0.5236~ & ~0.00432111~ & ~0.1193920~ & 410816 \\ 
115 & ~0.5406~ & ~0.00409845~ & ~0.1132400~ & 1036373 \\ 
120 & ~0.5560~ & ~0.00390982~ & ~0.1080280~ & 1080141 \\ 
125 & ~0.5700~ & ~0.00374705~ & ~0.1035310~ & 1500967 \\ 
130 & ~0.5829~ & ~0.00360381~ & ~0.0995733~ & 1887321 \\ 
135 & ~0.5947~ & ~0.00347548~ & ~0.0960274~ & 1216195 \\ 
140 & ~0.6056~ & ~0.00335869~ & ~0.0928007~ & 1912628 \\ 
145 & ~0.6158~ & ~0.00325107~ & ~0.0898270~ & 1383987 \\ 
150 & ~0.6252~ & ~0.00315088~ & ~0.0870590~ & 1338744 \\ 
155 & ~0.6341~ & ~0.00305689~ & ~0.0844619~ & 1005178 \\ 
160 & ~0.6425~ & ~0.00296817~ & ~0.0820105~ & 2215412 \\ 
165 & ~0.6504~ & ~0.00288403~ & ~0.0796857~ & 1596043 \\ 
170 & ~0.6579~ & ~0.00280394~ & ~0.0774727~ & 595253 \\ 
175 & ~0.6651~ & ~0.00272748~ & ~0.0753604~ & 1131649 \\ 
180 & ~0.6719~ & ~0.00265434~ & ~0.0733394~ & 1240884 \\ 
185 & ~0.6785~ & ~0.00258424~ & ~0.0714026~ & 436002 \\ 
190 & ~0.6848~ & ~0.00251696~ & ~0.0695436~ & 317895 \\ 
195 & ~0.6909~ & ~0.00245231~ & ~0.0677573~ & 361870 \\ 
200 & ~0.6968~ & ~0.00239013~ & ~0.0660393~ & 323968 \\ 
205 & ~0.7025~ & ~0.00233028~ & ~0.0643856~ & 158703 \\ 
210 & ~0.7080~ & ~0.00227263~ & ~0.0627928~ & 260064 \\ 
\hline
\end{tabular}

\caption{\label{tab:statistics}
Simulation parameters on the $16^3\times8$ lattice and the
number of configurations where the static quark potential's $\mu$-derivatives
have been computed.
The parameters are unchanged since Ref.~\cite{Borsanyi:2023wno}. The statistics
has been increased by one order of magnitude.
}
\end{table}

In this work we use 21 finite temperature ensembles
to calculate the $F_Q$ coefficients, as listed 
in Table.~\ref{tab:statistics}. For each configuration
we computed all eigenvalues of the reduced matrix for
both the light and the strange quark masses.
The first and the
last temperatures were actually excluded from the analysis.

In addition, four more ensembles were generated on
the same line of constant physics using $20^3\times10$,
$24^3\times12$, $28^3\times28$ and $32^3\times16$ for
the direct renormalization of the Polyakov loop. For
the $V_{Q\bar Q}$-based renormalization zero temperature
ensembles ($32^3\times64$ lattice) were generated for the
entire list with 1000 configurations in each.

\section{\label{app:formulas}Expansion formulas for the Polyakov loop}

With the notations $Q=|\avr{P}|^2$ and $Q^{(n)}=d^{n} Q/d\hat\mu^n$
we can convert the expansion of the norm of the Polyakov loop into
the static quark free energy:
\begin{eqnarray} 
F&=&-\frac{T}{2}\log{|\avr{P}|^2} = T \sum_{n=0,2,\dots} \frac{F_n}{n!} \left(\frac{\mu}{T}\right)^n \label{eq:FQ_sum} \\
F_0&=& -\frac{1}{2}\log(Q)\label{eq:FQ_0} \\
F_2 &=&-\frac{1}{2} \left[
+ \frac{Q^{(2)}}{Q}
\right]\label{eq:FQ_2} 
\\
F_4 &=&-\frac{1}{2}\left[
+ \frac{Q^{(4)}}{Q}
-3 \frac{\left(Q^{(2)}\right)^2}{Q^2}
\right]\label{eq:FQ_4} 
\\
F_6 &=&-\frac{1}{2}\left[
+ \frac{Q^{(6)}}{Q}
-15 \frac{Q^{(2)}Q^{(4)}}{Q^2}
+30 \frac{\left(Q^{(2)}\right)^3}{Q^3}
\right]\label{eq:FQ_6} 
\\
F_8 &=&
-\frac{1}{2}\left[
+ \frac{Q^{(8)}}{Q}
-35 \frac{\left(Q^{(4)}\right)^2}{Q^2}
-28 \frac{Q^{(2)}Q^{(6)}}{Q^2}\right.\nonumber\\
&&\qquad\left.
+420 \frac{\left(Q^{(2)}\right)^2Q^{(4)}}{Q^3}
-630 \frac{\left(Q^{(2)}\right)^4}{Q^4}
\right]\label{eq:FQ_8} 
\end{eqnarray}

In the following set of formulas we give the $\mu_u$ derivatives
of the norm ($Q$) of the Polyakov loop. These can be easily converted
into $\mu_B$ derivatives, if needed. For the full expansion all combinations
of $\mu_u$, $\mu_d$ and $\mu_s$ are needed, but we do not list them here. These
few, however, should make cross-checks of future implementations possible.
We remind the reader that the odd coefficients $A,C,\dots$ are imaginary.
Note that $P_I$ is not the imaginary part, but $i$ times the imaginary part of the
Polyakov loop.

\onecolumngrid

\begin{eqnarray*} 
\partial_u^2 Q&=&
+2 \langle P_R\rangle\langle B_u P_R\rangle
+2 \langle P_R\rangle\langle A_u A_u P_R\rangle
-2 \langle A_u P_I\rangle\langle A_u P_I\rangle
-2 \langle B_u\rangle\langle P_R\rangle\langle P_R\rangle
-2 \langle A_u A_u\rangle\langle P_R\rangle\langle P_R\rangle
\\
\partial_u^4 Q &=&
+2 \langle P_R\rangle\langle D_u P_R\rangle
+6 \langle P_R\rangle\langle B_u B_u P_R\rangle
+8 \langle P_R\rangle\langle A_u C_u P_R\rangle
+12 \langle P_R\rangle\langle A_u A_u B_u P_R\rangle
+2 \langle P_R\rangle\langle A_u A_u A_u A_u P_R\rangle
\nonumber\\
&&
+6 \langle B_u P_R\rangle\langle B_u P_R\rangle
-8 \langle A_u P_I\rangle\langle C_u P_I\rangle
-24 \langle A_u P_I\rangle\langle A_u B_u P_I\rangle
-8 \langle A_u P_I\rangle\langle A_u A_u A_u P_I\rangle
+12 \langle A_u A_u P_R\rangle\langle B_u P_R\rangle
\nonumber\\
&&
+6 \langle A_u A_u P_R\rangle\langle A_u A_u P_R\rangle
-2 \langle D_u\rangle\langle P_R\rangle\langle P_R\rangle
-24 \langle B_u\rangle\langle P_R\rangle\langle B_u P_R\rangle
-24 \langle B_u\rangle\langle P_R\rangle\langle A_u A_u P_R\rangle
\nonumber\\
&&
+24 \langle B_u\rangle\langle A_u P_I\rangle\langle A_u P_I\rangle
-6 \langle B_u B_u\rangle\langle P_R\rangle\langle P_R\rangle
-8 \langle A_u C_u\rangle\langle P_R\rangle\langle P_R\rangle
-24 \langle A_u A_u\rangle\langle P_R\rangle\langle B_u P_R\rangle
\nonumber\\
&&
-24 \langle A_u A_u\rangle\langle P_R\rangle\langle A_u A_u P_R\rangle
+24 \langle A_u A_u\rangle\langle A_u P_I\rangle\langle A_u P_I\rangle
-12 \langle A_u A_u B_u\rangle\langle P_R\rangle\langle P_R\rangle
-2 \langle A_u A_u A_u A_u\rangle\langle P_R\rangle\langle P_R\rangle
\nonumber\\
&&
+18 \langle B_u\rangle\langle B_u\rangle\langle P_R\rangle\langle P_R\rangle
+36 \langle A_u A_u\rangle\langle B_u\rangle\langle P_R\rangle\langle P_R\rangle
+18 \langle A_u A_u\rangle\langle A_u A_u\rangle\langle P_R\rangle\langle P_R\rangle
\end{eqnarray*}
\begin{eqnarray*} 
\partial_u^6 Q &=&
+2 \langle P_R\rangle\langle F_u P_R\rangle
+20 \langle P_R\rangle\langle C_u C_u P_R\rangle
+30 \langle P_R\rangle\langle B_u D_u P_R\rangle
+30 \langle P_R\rangle\langle B_u B_u B_u P_R\rangle
+12 \langle P_R\rangle\langle A_u E_u P_R\rangle
\nonumber\\
&&
+120 \langle P_R\rangle\langle A_u B_u C_u P_R\rangle
+30 \langle P_R\rangle\langle A_u A_u D_u P_R\rangle
+90 \langle P_R\rangle\langle A_u A_u B_u B_u P_R\rangle
+40 \langle P_R\rangle\langle A_u A_u A_u C_u P_R\rangle
\nonumber\\
&&
+30 \langle P_R\rangle\langle A_u A_u A_u A_u B_u P_R\rangle
+2 \langle P_R\rangle\langle A_u A_u A_u A_u A_u A_u P_R\rangle
-20 \langle C_u P_I\rangle\langle C_u P_I\rangle
+30 \langle B_u P_R\rangle\langle D_u P_R\rangle
\nonumber\\
&&
+90 \langle B_u P_R\rangle\langle B_u B_u P_R\rangle
+120 \langle B_u P_R\rangle\langle A_u C_u P_R\rangle
+180 \langle B_u P_R\rangle\langle A_u A_u B_u P_R\rangle
-12 \langle A_u P_I\rangle\langle E_u P_I\rangle
\nonumber\\
&&
-120 \langle A_u P_I\rangle\langle B_u C_u P_I\rangle
-60 \langle A_u P_I\rangle\langle A_u D_u P_I\rangle
-180 \langle A_u P_I\rangle\langle A_u B_u B_u P_I\rangle
-120 \langle A_u P_I\rangle\langle A_u A_u C_u P_I\rangle
\nonumber\\
&&
-120 \langle A_u P_I\rangle\langle A_u A_u A_u B_u P_I\rangle
-12 \langle A_u P_I\rangle\langle A_u A_u A_u A_u A_u P_I\rangle
-120 \langle A_u B_u P_I\rangle\langle C_u P_I\rangle
-180 \langle A_u B_u P_I\rangle\langle A_u B_u P_I\rangle
\nonumber\\
&&
+30 \langle A_u A_u P_R\rangle\langle D_u P_R\rangle
+90 \langle A_u A_u P_R\rangle\langle B_u B_u P_R\rangle
+120 \langle A_u A_u P_R\rangle\langle A_u C_u P_R\rangle
+180 \langle A_u A_u P_R\rangle\langle A_u A_u B_u P_R\rangle
\nonumber\\
&&
+30 \langle A_u A_u P_R\rangle\langle A_u A_u A_u A_u P_R\rangle
-40 \langle A_u A_u A_u P_I\rangle\langle C_u P_I\rangle
-120 \langle A_u A_u A_u P_I\rangle\langle A_u B_u P_I\rangle
-20 \langle A_u A_u A_u P_I\rangle\langle A_u A_u A_u P_I\rangle
\nonumber\\
&&
+30 \langle A_u A_u A_u A_u P_R\rangle\langle B_u P_R\rangle
-2 \langle F_u\rangle\langle P_R\rangle\langle P_R\rangle
-60 \langle D_u\rangle\langle P_R\rangle\langle B_u P_R\rangle
-60 \langle D_u\rangle\langle P_R\rangle\langle A_u A_u P_R\rangle
\nonumber\\
&&
+60 \langle D_u\rangle\langle A_u P_I\rangle\langle A_u P_I\rangle
-20 \langle C_u C_u\rangle\langle P_R\rangle\langle P_R\rangle
-60 \langle B_u\rangle\langle P_R\rangle\langle D_u P_R\rangle
-180 \langle B_u\rangle\langle P_R\rangle\langle B_u B_u P_R\rangle
\nonumber\\
&&
-240 \langle B_u\rangle\langle P_R\rangle\langle A_u C_u P_R\rangle
-360 \langle B_u\rangle\langle P_R\rangle\langle A_u A_u B_u P_R\rangle
-60 \langle B_u\rangle\langle P_R\rangle\langle A_u A_u A_u A_u P_R\rangle
-180 \langle B_u\rangle\langle B_u P_R\rangle\langle B_u P_R\rangle
\nonumber\\
&&
+240 \langle B_u\rangle\langle A_u P_I\rangle\langle C_u P_I\rangle
+720 \langle B_u\rangle\langle A_u P_I\rangle\langle A_u B_u P_I\rangle
+240 \langle B_u\rangle\langle A_u P_I\rangle\langle A_u A_u A_u P_I\rangle
-360 \langle B_u\rangle\langle A_u A_u P_R\rangle\langle B_u P_R\rangle
\nonumber\\
&&
-180 \langle B_u\rangle\langle A_u A_u P_R\rangle\langle A_u A_u P_R\rangle
-30 \langle B_u D_u\rangle\langle P_R\rangle\langle P_R\rangle
-180 \langle B_u B_u\rangle\langle P_R\rangle\langle B_u P_R\rangle
-180 \langle B_u B_u\rangle\langle P_R\rangle\langle A_u A_u P_R\rangle
\nonumber\\
&&
+180 \langle B_u B_u\rangle\langle A_u P_I\rangle\langle A_u P_I\rangle
-30 \langle B_u B_u B_u\rangle\langle P_R\rangle\langle P_R\rangle
-12 \langle A_u E_u\rangle\langle P_R\rangle\langle P_R\rangle
-240 \langle A_u C_u\rangle\langle P_R\rangle\langle B_u P_R\rangle
\nonumber\\
&&
-240 \langle A_u C_u\rangle\langle P_R\rangle\langle A_u A_u P_R\rangle
+240 \langle A_u C_u\rangle\langle A_u P_I\rangle\langle A_u P_I\rangle
-120 \langle A_u B_u C_u\rangle\langle P_R\rangle\langle P_R\rangle
-60 \langle A_u A_u\rangle\langle P_R\rangle\langle D_u P_R\rangle
\nonumber\\
&&
-180 \langle A_u A_u\rangle\langle P_R\rangle\langle B_u B_u P_R\rangle
-240 \langle A_u A_u\rangle\langle P_R\rangle\langle A_u C_u P_R\rangle
-360 \langle A_u A_u\rangle\langle P_R\rangle\langle A_u A_u B_u P_R\rangle
\nonumber\\
&&
-60 \langle A_u A_u\rangle\langle P_R\rangle\langle A_u A_u A_u A_u P_R\rangle
-180 \langle A_u A_u\rangle\langle B_u P_R\rangle\langle B_u P_R\rangle
+240 \langle A_u A_u\rangle\langle A_u P_I\rangle\langle C_u P_I\rangle
\nonumber\\
&&
+720 \langle A_u A_u\rangle\langle A_u P_I\rangle\langle A_u B_u P_I\rangle
+240 \langle A_u A_u\rangle\langle A_u P_I\rangle\langle A_u A_u A_u P_I\rangle
-360 \langle A_u A_u\rangle\langle A_u A_u P_R\rangle\langle B_u P_R\rangle
\nonumber\\
&&
-180 \langle A_u A_u\rangle\langle A_u A_u P_R\rangle\langle A_u A_u P_R\rangle
-30 \langle A_u A_u D_u\rangle\langle P_R\rangle\langle P_R\rangle
-360 \langle A_u A_u B_u\rangle\langle P_R\rangle\langle B_u P_R\rangle
\nonumber\\
&&
-360 \langle A_u A_u B_u\rangle\langle P_R\rangle\langle A_u A_u P_R\rangle
+360 \langle A_u A_u B_u\rangle\langle A_u P_I\rangle\langle A_u P_I\rangle
-90 \langle A_u A_u B_u B_u\rangle\langle P_R\rangle\langle P_R\rangle
\nonumber\\
&&
-40 \langle A_u A_u A_u C_u\rangle\langle P_R\rangle\langle P_R\rangle
-60 \langle A_u A_u A_u A_u\rangle\langle P_R\rangle\langle B_u P_R\rangle
-60 \langle A_u A_u A_u A_u\rangle\langle P_R\rangle\langle A_u A_u P_R\rangle
\nonumber\\
&&
+60 \langle A_u A_u A_u A_u\rangle\langle A_u P_I\rangle\langle A_u P_I\rangle
-30 \langle A_u A_u A_u A_u B_u\rangle\langle P_R\rangle\langle P_R\rangle
-2 \langle A_u A_u A_u A_u A_u A_u\rangle\langle P_R\rangle\langle P_R\rangle
\nonumber\\
&&
+90 \langle B_u\rangle\langle D_u\rangle\langle P_R\rangle\langle P_R\rangle
+540 \langle B_u\rangle\langle B_u\rangle\langle P_R\rangle\langle B_u P_R\rangle
+540 \langle B_u\rangle\langle B_u\rangle\langle P_R\rangle\langle A_u A_u P_R\rangle
\nonumber\\
&&
-540 \langle B_u\rangle\langle B_u\rangle\langle A_u P_I\rangle\langle A_u P_I\rangle
+270 \langle B_u\rangle\langle B_u B_u\rangle\langle P_R\rangle\langle P_R\rangle
+360 \langle B_u\rangle\langle A_u C_u\rangle\langle P_R\rangle\langle P_R\rangle
\nonumber\\
&&
+540 \langle B_u\rangle\langle A_u A_u B_u\rangle\langle P_R\rangle\langle P_R\rangle
+90 \langle A_u A_u\rangle\langle D_u\rangle\langle P_R\rangle\langle P_R\rangle
+1080 \langle A_u A_u\rangle\langle B_u\rangle\langle P_R\rangle\langle B_u P_R\rangle
\nonumber\\
&&
+1080 \langle A_u A_u\rangle\langle B_u\rangle\langle P_R\rangle\langle A_u A_u P_R\rangle
-1080 \langle A_u A_u\rangle\langle B_u\rangle\langle A_u P_I\rangle\langle A_u P_I\rangle
+270 \langle A_u A_u\rangle\langle B_u B_u\rangle\langle P_R\rangle\langle P_R\rangle
\nonumber\\
&&
+360 \langle A_u A_u\rangle\langle A_u C_u\rangle\langle P_R\rangle\langle P_R\rangle
+540 \langle A_u A_u\rangle\langle A_u A_u\rangle\langle P_R\rangle\langle B_u P_R\rangle
+540 \langle A_u A_u\rangle\langle A_u A_u\rangle\langle P_R\rangle\langle A_u A_u P_R\rangle
\nonumber\\
&&
-540 \langle A_u A_u\rangle\langle A_u A_u\rangle\langle A_u P_I\rangle\langle A_u P_I\rangle
+540 \langle A_u A_u\rangle\langle A_u A_u B_u\rangle\langle P_R\rangle\langle P_R\rangle
+90 \langle A_u A_u\rangle\langle A_u A_u A_u A_u\rangle\langle P_R\rangle\langle P_R\rangle
\nonumber\\
&&
+90 \langle A_u A_u A_u A_u\rangle\langle B_u\rangle\langle P_R\rangle\langle P_R\rangle
-360 \langle B_u\rangle\langle B_u\rangle\langle B_u\rangle\langle P_R\rangle\langle P_R\rangle
-1080 \langle A_u A_u\rangle\langle B_u\rangle\langle B_u\rangle\langle P_R\rangle\langle P_R\rangle
\nonumber\\
&&
-1080 \langle A_u A_u\rangle\langle A_u A_u\rangle\langle B_u\rangle\langle P_R\rangle\langle P_R\rangle
-360 \langle A_u A_u\rangle\langle A_u A_u\rangle\langle A_u A_u\rangle\langle P_R\rangle\langle P_R\rangle\\
\partial_u^8 Q &=& \textrm{has~405~terms}
\end{eqnarray*}

\twocolumngrid
\bibliography{thermo}

\begin{thebibliography}{68}%
\makeatletter
\providecommand \@ifxundefined [1]{%
 \@ifx{#1\undefined}
}%
\providecommand \@ifnum [1]{%
 \ifnum #1\expandafter \@firstoftwo
 \else \expandafter \@secondoftwo
 \fi
}%
\providecommand \@ifx [1]{%
 \ifx #1\expandafter \@firstoftwo
 \else \expandafter \@secondoftwo
 \fi
}%
\providecommand \natexlab [1]{#1}%
\providecommand \enquote  [1]{``#1''}%
\providecommand \bibnamefont  [1]{#1}%
\providecommand \bibfnamefont [1]{#1}%
\providecommand \citenamefont [1]{#1}%
\providecommand \href@noop [0]{\@secondoftwo}%
\providecommand \href [0]{\begingroup \@sanitize@url \@href}%
\providecommand \@href[1]{\@@startlink{#1}\@@href}%
\providecommand \@@href[1]{\endgroup#1\@@endlink}%
\providecommand \@sanitize@url [0]{\catcode `\\12\catcode `\$12\catcode `\&12\catcode `\#12\catcode `\^12\catcode `\_12\catcode `\%12\relax}%
\providecommand \@@startlink[1]{}%
\providecommand \@@endlink[0]{}%
\providecommand \url  [0]{\begingroup\@sanitize@url \@url }%
\providecommand \@url [1]{\endgroup\@href {#1}{\urlprefix }}%
\providecommand \urlprefix  [0]{URL }%
\providecommand \Eprint [0]{\href }%
\providecommand \doibase [0]{http://dx.doi.org/}%
\providecommand \selectlanguage [0]{\@gobble}%
\providecommand \bibinfo  [0]{\@secondoftwo}%
\providecommand \bibfield  [0]{\@secondoftwo}%
\providecommand \translation [1]{[#1]}%
\providecommand \BibitemOpen [0]{}%
\providecommand \bibitemStop [0]{}%
\providecommand \bibitemNoStop [0]{.\EOS\space}%
\providecommand \EOS [0]{\spacefactor3000\relax}%
\providecommand \BibitemShut  [1]{\csname bibitem#1\endcsname}%
\let\auto@bib@innerbib\@empty
\bibitem [{\citenamefont {Aoki}\ \emph {et~al.}(2006{\natexlab{a}})\citenamefont {Aoki}, \citenamefont {Endrodi}, \citenamefont {Fodor}, \citenamefont {Katz},\ and\ \citenamefont {Szabo}}]{Aoki:2006we}%
  \BibitemOpen
  \bibfield  {author} {\bibinfo {author} {\bibfnamefont {Y.}~\bibnamefont {Aoki}}, \bibinfo {author} {\bibfnamefont {G.}~\bibnamefont {Endrodi}}, \bibinfo {author} {\bibfnamefont {Z.}~\bibnamefont {Fodor}}, \bibinfo {author} {\bibfnamefont {S.}~\bibnamefont {Katz}}, \ and\ \bibinfo {author} {\bibfnamefont {K.}~\bibnamefont {Szabo}},\ }\href {\doibase 10.1038/nature05120} {\bibfield  {journal} {\bibinfo  {journal} {Nature}\ }\textbf {\bibinfo {volume} {443}},\ \bibinfo {pages} {675} (\bibinfo {year} {2006}{\natexlab{a}})},\ \Eprint {http://arxiv.org/abs/hep-lat/0611014} {arXiv:hep-lat/0611014 [hep-lat]} \BibitemShut {NoStop}%
\bibitem [{\citenamefont {Borsanyi}\ \emph {et~al.}(2010)\citenamefont {Borsanyi} \emph {et~al.}}]{Borsanyi:2010bp}%
  \BibitemOpen
  \bibfield  {author} {\bibinfo {author} {\bibfnamefont {S.}~\bibnamefont {Borsanyi}} \emph {et~al.} (\bibinfo {collaboration} {Wuppertal-Budapest Collaboration}),\ }\href {\doibase 10.1007/JHEP09(2010)073} {\bibfield  {journal} {\bibinfo  {journal} {JHEP}\ }\textbf {\bibinfo {volume} {1009}},\ \bibinfo {pages} {073} (\bibinfo {year} {2010})},\ \Eprint {http://arxiv.org/abs/1005.3508} {arXiv:1005.3508 [hep-lat]} \BibitemShut {NoStop}%
\bibitem [{\citenamefont {Ding}\ \emph {et~al.}(2019{\natexlab{a}})\citenamefont {Ding} \emph {et~al.}}]{HotQCD:2019xnw}%
  \BibitemOpen
  \bibfield  {author} {\bibinfo {author} {\bibfnamefont {H.~T.}\ \bibnamefont {Ding}} \emph {et~al.} (\bibinfo {collaboration} {HotQCD}),\ }\href {\doibase 10.1103/PhysRevLett.123.062002} {\bibfield  {journal} {\bibinfo  {journal} {Phys. Rev. Lett.}\ }\textbf {\bibinfo {volume} {123}},\ \bibinfo {pages} {062002} (\bibinfo {year} {2019}{\natexlab{a}})},\ \Eprint {http://arxiv.org/abs/1903.04801} {arXiv:1903.04801 [hep-lat]} \BibitemShut {NoStop}%
\bibitem [{\citenamefont {Bazavov}\ \emph {et~al.}(2012)\citenamefont {Bazavov}, \citenamefont {Bhattacharya}, \citenamefont {Cheng}, \citenamefont {DeTar}, \citenamefont {Ding} \emph {et~al.}}]{Bazavov:2011nk}%
  \BibitemOpen
  \bibfield  {author} {\bibinfo {author} {\bibfnamefont {A.}~\bibnamefont {Bazavov}}, \bibinfo {author} {\bibfnamefont {T.}~\bibnamefont {Bhattacharya}}, \bibinfo {author} {\bibfnamefont {M.}~\bibnamefont {Cheng}}, \bibinfo {author} {\bibfnamefont {C.}~\bibnamefont {DeTar}}, \bibinfo {author} {\bibfnamefont {H.}~\bibnamefont {Ding}},  \emph {et~al.},\ }\href {\doibase 10.1103/PhysRevD.85.054503} {\bibfield  {journal} {\bibinfo  {journal} {Phys.Rev.}\ }\textbf {\bibinfo {volume} {D85}},\ \bibinfo {pages} {054503} (\bibinfo {year} {2012})},\ \Eprint {http://arxiv.org/abs/1111.1710} {arXiv:1111.1710 [hep-lat]} \BibitemShut {NoStop}%
\bibitem [{\citenamefont {Kotov}\ \emph {et~al.}(2021)\citenamefont {Kotov}, \citenamefont {Lombardo},\ and\ \citenamefont {Trunin}}]{Kotov:2021rah}%
  \BibitemOpen
  \bibfield  {author} {\bibinfo {author} {\bibfnamefont {A.~Y.}\ \bibnamefont {Kotov}}, \bibinfo {author} {\bibfnamefont {M.~P.}\ \bibnamefont {Lombardo}}, \ and\ \bibinfo {author} {\bibfnamefont {A.}~\bibnamefont {Trunin}},\ }\href {\doibase 10.1016/j.physletb.2021.136749} {\bibfield  {journal} {\bibinfo  {journal} {Phys. Lett. B}\ }\textbf {\bibinfo {volume} {823}},\ \bibinfo {pages} {136749} (\bibinfo {year} {2021})},\ \Eprint {http://arxiv.org/abs/2105.09842} {arXiv:2105.09842 [hep-lat]} \BibitemShut {NoStop}%
\bibitem [{\citenamefont {Cuteri}\ \emph {et~al.}(2021{\natexlab{a}})\citenamefont {Cuteri}, \citenamefont {Philipsen},\ and\ \citenamefont {Sciarra}}]{Cuteri:2021ikv}%
  \BibitemOpen
  \bibfield  {author} {\bibinfo {author} {\bibfnamefont {F.}~\bibnamefont {Cuteri}}, \bibinfo {author} {\bibfnamefont {O.}~\bibnamefont {Philipsen}}, \ and\ \bibinfo {author} {\bibfnamefont {A.}~\bibnamefont {Sciarra}},\ }\href {\doibase 10.1007/JHEP11(2021)141} {\bibfield  {journal} {\bibinfo  {journal} {JHEP}\ }\textbf {\bibinfo {volume} {11}},\ \bibinfo {pages} {141} (\bibinfo {year} {2021}{\natexlab{a}})},\ \Eprint {http://arxiv.org/abs/2107.12739} {arXiv:2107.12739 [hep-lat]} \BibitemShut {NoStop}%
\bibitem [{\citenamefont {Pratt}\ \emph {et~al.}(2015)\citenamefont {Pratt}, \citenamefont {Sangaline}, \citenamefont {Sorensen},\ and\ \citenamefont {Wang}}]{Pratt:2015zsa}%
  \BibitemOpen
  \bibfield  {author} {\bibinfo {author} {\bibfnamefont {S.}~\bibnamefont {Pratt}}, \bibinfo {author} {\bibfnamefont {E.}~\bibnamefont {Sangaline}}, \bibinfo {author} {\bibfnamefont {P.}~\bibnamefont {Sorensen}}, \ and\ \bibinfo {author} {\bibfnamefont {H.}~\bibnamefont {Wang}},\ }\href {\doibase 10.1103/PhysRevLett.114.202301} {\bibfield  {journal} {\bibinfo  {journal} {Phys. Rev. Lett.}\ }\textbf {\bibinfo {volume} {114}},\ \bibinfo {pages} {202301} (\bibinfo {year} {2015})},\ \Eprint {http://arxiv.org/abs/1501.04042} {arXiv:1501.04042 [nucl-th]} \BibitemShut {NoStop}%
\bibitem [{\citenamefont {Pang}\ \emph {et~al.}(2018)\citenamefont {Pang}, \citenamefont {Zhou}, \citenamefont {Su}, \citenamefont {Petersen}, \citenamefont {St\"ocker},\ and\ \citenamefont {Wang}}]{Pang:2016vdc}%
  \BibitemOpen
  \bibfield  {author} {\bibinfo {author} {\bibfnamefont {L.-G.}\ \bibnamefont {Pang}}, \bibinfo {author} {\bibfnamefont {K.}~\bibnamefont {Zhou}}, \bibinfo {author} {\bibfnamefont {N.}~\bibnamefont {Su}}, \bibinfo {author} {\bibfnamefont {H.}~\bibnamefont {Petersen}}, \bibinfo {author} {\bibfnamefont {H.}~\bibnamefont {St\"ocker}}, \ and\ \bibinfo {author} {\bibfnamefont {X.-N.}\ \bibnamefont {Wang}},\ }\href {\doibase 10.1038/s41467-017-02726-3} {\bibfield  {journal} {\bibinfo  {journal} {Nature Commun.}\ }\textbf {\bibinfo {volume} {9}},\ \bibinfo {pages} {210} (\bibinfo {year} {2018})},\ \Eprint {http://arxiv.org/abs/1612.04262} {arXiv:1612.04262 [hep-ph]} \BibitemShut {NoStop}%
\bibitem [{\citenamefont {Abdallah}\ \emph {et~al.}(2021)\citenamefont {Abdallah} \emph {et~al.}}]{STAR:2021rls}%
  \BibitemOpen
  \bibfield  {author} {\bibinfo {author} {\bibfnamefont {M.}~\bibnamefont {Abdallah}} \emph {et~al.} (\bibinfo {collaboration} {STAR}),\ }\href {\doibase 10.1103/PhysRevLett.127.262301} {\bibfield  {journal} {\bibinfo  {journal} {Phys. Rev. Lett.}\ }\textbf {\bibinfo {volume} {127}},\ \bibinfo {pages} {262301} (\bibinfo {year} {2021})},\ \Eprint {http://arxiv.org/abs/2105.14698} {arXiv:2105.14698 [nucl-ex]} \BibitemShut {NoStop}%
\bibitem [{\citenamefont {Pisarski}\ and\ \citenamefont {Wilczek}(1984)}]{Pisarski:1983ms}%
  \BibitemOpen
  \bibfield  {author} {\bibinfo {author} {\bibfnamefont {R.~D.}\ \bibnamefont {Pisarski}}\ and\ \bibinfo {author} {\bibfnamefont {F.}~\bibnamefont {Wilczek}},\ }\href {\doibase 10.1103/PhysRevD.29.338} {\bibfield  {journal} {\bibinfo  {journal} {Phys.Rev.}\ }\textbf {\bibinfo {volume} {D29}},\ \bibinfo {pages} {338} (\bibinfo {year} {1984})}\BibitemShut {NoStop}%
\bibitem [{\citenamefont {Pelissetto}\ and\ \citenamefont {Vicari}(2013)}]{Pelissetto:2013hqa}%
  \BibitemOpen
  \bibfield  {author} {\bibinfo {author} {\bibfnamefont {A.}~\bibnamefont {Pelissetto}}\ and\ \bibinfo {author} {\bibfnamefont {E.}~\bibnamefont {Vicari}},\ }\href {\doibase 10.1103/PhysRevD.88.105018} {\bibfield  {journal} {\bibinfo  {journal} {Phys.Rev.}\ }\textbf {\bibinfo {volume} {D88}},\ \bibinfo {pages} {105018} (\bibinfo {year} {2013})},\ \Eprint {http://arxiv.org/abs/1309.5446} {arXiv:1309.5446 [hep-lat]} \BibitemShut {NoStop}%
\bibitem [{\citenamefont {Fejos}\ and\ \citenamefont {Hatsuda}(2024)}]{Fejos:2024bgl}%
  \BibitemOpen
  \bibfield  {author} {\bibinfo {author} {\bibfnamefont {G.}~\bibnamefont {Fejos}}\ and\ \bibinfo {author} {\bibfnamefont {T.}~\bibnamefont {Hatsuda}},\ }\href {\doibase 10.1103/PhysRevD.110.016021} {\bibfield  {journal} {\bibinfo  {journal} {Phys. Rev. D}\ }\textbf {\bibinfo {volume} {110}},\ \bibinfo {pages} {016021} (\bibinfo {year} {2024})},\ \Eprint {http://arxiv.org/abs/2404.00554} {arXiv:2404.00554 [hep-ph]} \BibitemShut {NoStop}%
\bibitem [{\citenamefont {Aoki}\ \emph {et~al.}(2006{\natexlab{b}})\citenamefont {Aoki}, \citenamefont {Fodor}, \citenamefont {Katz},\ and\ \citenamefont {Szabo}}]{Aoki:2006br}%
  \BibitemOpen
  \bibfield  {author} {\bibinfo {author} {\bibfnamefont {Y.}~\bibnamefont {Aoki}}, \bibinfo {author} {\bibfnamefont {Z.}~\bibnamefont {Fodor}}, \bibinfo {author} {\bibfnamefont {S.}~\bibnamefont {Katz}}, \ and\ \bibinfo {author} {\bibfnamefont {K.}~\bibnamefont {Szabo}},\ }\href {\doibase 10.1016/j.physletb.2006.10.021} {\bibfield  {journal} {\bibinfo  {journal} {Phys.Lett.}\ }\textbf {\bibinfo {volume} {B643}},\ \bibinfo {pages} {46} (\bibinfo {year} {2006}{\natexlab{b}})},\ \Eprint {http://arxiv.org/abs/hep-lat/0609068} {arXiv:hep-lat/0609068 [hep-lat]} \BibitemShut {NoStop}%
\bibitem [{\citenamefont {Aoki}\ \emph {et~al.}(2009)\citenamefont {Aoki}, \citenamefont {Borsanyi}, \citenamefont {Durr}, \citenamefont {Fodor}, \citenamefont {Katz} \emph {et~al.}}]{Aoki:2009sc}%
  \BibitemOpen
  \bibfield  {author} {\bibinfo {author} {\bibfnamefont {Y.}~\bibnamefont {Aoki}}, \bibinfo {author} {\bibfnamefont {S.}~\bibnamefont {Borsanyi}}, \bibinfo {author} {\bibfnamefont {S.}~\bibnamefont {Durr}}, \bibinfo {author} {\bibfnamefont {Z.}~\bibnamefont {Fodor}}, \bibinfo {author} {\bibfnamefont {S.~D.}\ \bibnamefont {Katz}},  \emph {et~al.},\ }\href {\doibase 10.1088/1126-6708/2009/06/088} {\bibfield  {journal} {\bibinfo  {journal} {JHEP}\ }\textbf {\bibinfo {volume} {0906}},\ \bibinfo {pages} {088} (\bibinfo {year} {2009})},\ \Eprint {http://arxiv.org/abs/0903.4155} {arXiv:0903.4155 [hep-lat]} \BibitemShut {NoStop}%
\bibitem [{\citenamefont {Ejiri}\ \emph {et~al.}(2009)\citenamefont {Ejiri}, \citenamefont {Karsch}, \citenamefont {Laermann}, \citenamefont {Miao}, \citenamefont {Mukherjee} \emph {et~al.}}]{Ejiri:2009ac}%
  \BibitemOpen
  \bibfield  {author} {\bibinfo {author} {\bibfnamefont {S.}~\bibnamefont {Ejiri}}, \bibinfo {author} {\bibfnamefont {F.}~\bibnamefont {Karsch}}, \bibinfo {author} {\bibfnamefont {E.}~\bibnamefont {Laermann}}, \bibinfo {author} {\bibfnamefont {C.}~\bibnamefont {Miao}}, \bibinfo {author} {\bibfnamefont {S.}~\bibnamefont {Mukherjee}},  \emph {et~al.},\ }\href {\doibase 10.1103/PhysRevD.80.094505} {\bibfield  {journal} {\bibinfo  {journal} {Phys.Rev.}\ }\textbf {\bibinfo {volume} {D80}},\ \bibinfo {pages} {094505} (\bibinfo {year} {2009})},\ \Eprint {http://arxiv.org/abs/0909.5122} {arXiv:0909.5122 [hep-lat]} \BibitemShut {NoStop}%
\bibitem [{\citenamefont {Ding}\ \emph {et~al.}(2019{\natexlab{b}})\citenamefont {Ding} \emph {et~al.}}]{Ding:2019prx}%
  \BibitemOpen
  \bibfield  {author} {\bibinfo {author} {\bibfnamefont {H.~T.}\ \bibnamefont {Ding}} \emph {et~al.},\ }\href {\doibase 10.1103/PhysRevLett.123.062002} {\bibfield  {journal} {\bibinfo  {journal} {Phys. Rev. Lett.}\ }\textbf {\bibinfo {volume} {123}},\ \bibinfo {pages} {062002} (\bibinfo {year} {2019}{\natexlab{b}})},\ \Eprint {http://arxiv.org/abs/1903.04801} {arXiv:1903.04801 [hep-lat]} \BibitemShut {NoStop}%
\bibitem [{\citenamefont {Ding}\ \emph {et~al.}(2024)\citenamefont {Ding}, \citenamefont {Kaczmarek}, \citenamefont {Karsch}, \citenamefont {Petreczky}, \citenamefont {Sarkar}, \citenamefont {Schmidt},\ and\ \citenamefont {Sharma}}]{Ding:2024sux}%
  \BibitemOpen
  \bibfield  {author} {\bibinfo {author} {\bibfnamefont {H.~T.}\ \bibnamefont {Ding}}, \bibinfo {author} {\bibfnamefont {O.}~\bibnamefont {Kaczmarek}}, \bibinfo {author} {\bibfnamefont {F.}~\bibnamefont {Karsch}}, \bibinfo {author} {\bibfnamefont {P.}~\bibnamefont {Petreczky}}, \bibinfo {author} {\bibfnamefont {M.}~\bibnamefont {Sarkar}}, \bibinfo {author} {\bibfnamefont {C.}~\bibnamefont {Schmidt}}, \ and\ \bibinfo {author} {\bibfnamefont {S.}~\bibnamefont {Sharma}},\ }\href {\doibase 10.1103/PhysRevD.109.114516} {\bibfield  {journal} {\bibinfo  {journal} {Phys. Rev. D}\ }\textbf {\bibinfo {volume} {109}},\ \bibinfo {pages} {114516} (\bibinfo {year} {2024})},\ \Eprint {http://arxiv.org/abs/2403.09390} {arXiv:2403.09390 [hep-lat]} \BibitemShut {NoStop}%
\bibitem [{\citenamefont {Borsanyi}\ \emph {et~al.}(2020)\citenamefont {Borsanyi}, \citenamefont {Fodor}, \citenamefont {Guenther}, \citenamefont {Kara}, \citenamefont {Katz}, \citenamefont {Parotto}, \citenamefont {Pasztor}, \citenamefont {Ratti},\ and\ \citenamefont {Szabo}}]{Borsanyi:2020fev}%
  \BibitemOpen
  \bibfield  {author} {\bibinfo {author} {\bibfnamefont {S.}~\bibnamefont {Borsanyi}}, \bibinfo {author} {\bibfnamefont {Z.}~\bibnamefont {Fodor}}, \bibinfo {author} {\bibfnamefont {J.~N.}\ \bibnamefont {Guenther}}, \bibinfo {author} {\bibfnamefont {R.}~\bibnamefont {Kara}}, \bibinfo {author} {\bibfnamefont {S.~D.}\ \bibnamefont {Katz}}, \bibinfo {author} {\bibfnamefont {P.}~\bibnamefont {Parotto}}, \bibinfo {author} {\bibfnamefont {A.}~\bibnamefont {Pasztor}}, \bibinfo {author} {\bibfnamefont {C.}~\bibnamefont {Ratti}}, \ and\ \bibinfo {author} {\bibfnamefont {K.~K.}\ \bibnamefont {Szabo}},\ }\href@noop {} {\bibfield  {journal} {\bibinfo  {journal} {Phys. Rev. Lett.}\ }\textbf {\bibinfo {volume} {125}},\ \bibinfo {pages} {052001} (\bibinfo {year} {2020})},\ \Eprint {http://arxiv.org/abs/2002.02821} {arXiv:2002.02821 [hep-lat]} \BibitemShut {NoStop}%
\bibitem [{\citenamefont {Gunkel}\ and\ \citenamefont {Fischer}(2021)}]{Gunkel:2021oya}%
  \BibitemOpen
  \bibfield  {author} {\bibinfo {author} {\bibfnamefont {P.~J.}\ \bibnamefont {Gunkel}}\ and\ \bibinfo {author} {\bibfnamefont {C.~S.}\ \bibnamefont {Fischer}},\ }\href {\doibase 10.1103/PhysRevD.104.054022} {\bibfield  {journal} {\bibinfo  {journal} {Phys. Rev. D}\ }\textbf {\bibinfo {volume} {104}},\ \bibinfo {pages} {054022} (\bibinfo {year} {2021})},\ \Eprint {http://arxiv.org/abs/2106.08356} {arXiv:2106.08356 [hep-ph]} \BibitemShut {NoStop}%
\bibitem [{\citenamefont {Adamczyk}\ \emph {et~al.}(2017)\citenamefont {Adamczyk} \emph {et~al.}}]{STAR:2017sal}%
  \BibitemOpen
  \bibfield  {author} {\bibinfo {author} {\bibfnamefont {L.}~\bibnamefont {Adamczyk}} \emph {et~al.} (\bibinfo {collaboration} {STAR}),\ }\href {\doibase 10.1103/PhysRevC.96.044904} {\bibfield  {journal} {\bibinfo  {journal} {Phys. Rev. C}\ }\textbf {\bibinfo {volume} {96}},\ \bibinfo {pages} {044904} (\bibinfo {year} {2017})},\ \Eprint {http://arxiv.org/abs/1701.07065} {arXiv:1701.07065 [nucl-ex]} \BibitemShut {NoStop}%
\bibitem [{\citenamefont {Andronic}\ \emph {et~al.}(2018)\citenamefont {Andronic}, \citenamefont {Braun-Munzinger}, \citenamefont {Redlich},\ and\ \citenamefont {Stachel}}]{Andronic:2017pug}%
  \BibitemOpen
  \bibfield  {author} {\bibinfo {author} {\bibfnamefont {A.}~\bibnamefont {Andronic}}, \bibinfo {author} {\bibfnamefont {P.}~\bibnamefont {Braun-Munzinger}}, \bibinfo {author} {\bibfnamefont {K.}~\bibnamefont {Redlich}}, \ and\ \bibinfo {author} {\bibfnamefont {J.}~\bibnamefont {Stachel}},\ }\href {\doibase 10.1038/s41586-018-0491-6} {\bibfield  {journal} {\bibinfo  {journal} {Nature}\ }\textbf {\bibinfo {volume} {561}},\ \bibinfo {pages} {321} (\bibinfo {year} {2018})},\ \Eprint {http://arxiv.org/abs/1710.09425} {arXiv:1710.09425 [nucl-th]} \BibitemShut {NoStop}%
\bibitem [{\citenamefont {Borsanyi}\ \emph {et~al.}(2022)\citenamefont {Borsanyi}, \citenamefont {R.}, \citenamefont {Fodor}, \citenamefont {Godzieba}, \citenamefont {Parotto},\ and\ \citenamefont {Sexty}}]{Borsanyi:2022xml}%
  \BibitemOpen
  \bibfield  {author} {\bibinfo {author} {\bibfnamefont {S.}~\bibnamefont {Borsanyi}}, \bibinfo {author} {\bibfnamefont {K.}~\bibnamefont {R.}}, \bibinfo {author} {\bibfnamefont {Z.}~\bibnamefont {Fodor}}, \bibinfo {author} {\bibfnamefont {D.~A.}\ \bibnamefont {Godzieba}}, \bibinfo {author} {\bibfnamefont {P.}~\bibnamefont {Parotto}}, \ and\ \bibinfo {author} {\bibfnamefont {D.}~\bibnamefont {Sexty}},\ }\href {\doibase 10.1103/PhysRevD.105.074513} {\bibfield  {journal} {\bibinfo  {journal} {Phys. Rev. D}\ }\textbf {\bibinfo {volume} {105}},\ \bibinfo {pages} {074513} (\bibinfo {year} {2022})},\ \Eprint {http://arxiv.org/abs/2202.05234} {arXiv:2202.05234 [hep-lat]} \BibitemShut {NoStop}%
\bibitem [{\citenamefont {McLerran}\ and\ \citenamefont {Svetitsky}(1981)}]{McLerran:1981pb}%
  \BibitemOpen
  \bibfield  {author} {\bibinfo {author} {\bibfnamefont {L.~D.}\ \bibnamefont {McLerran}}\ and\ \bibinfo {author} {\bibfnamefont {B.}~\bibnamefont {Svetitsky}},\ }\href {\doibase 10.1103/PhysRevD.24.450} {\bibfield  {journal} {\bibinfo  {journal} {Phys. Rev. D}\ }\textbf {\bibinfo {volume} {24}},\ \bibinfo {pages} {450} (\bibinfo {year} {1981})}\BibitemShut {NoStop}%
\bibitem [{\citenamefont {Cuteri}\ \emph {et~al.}(2021{\natexlab{b}})\citenamefont {Cuteri}, \citenamefont {Philipsen}, \citenamefont {Sch\"on},\ and\ \citenamefont {Sciarra}}]{Cuteri:2020yke}%
  \BibitemOpen
  \bibfield  {author} {\bibinfo {author} {\bibfnamefont {F.}~\bibnamefont {Cuteri}}, \bibinfo {author} {\bibfnamefont {O.}~\bibnamefont {Philipsen}}, \bibinfo {author} {\bibfnamefont {A.}~\bibnamefont {Sch\"on}}, \ and\ \bibinfo {author} {\bibfnamefont {A.}~\bibnamefont {Sciarra}},\ }\href {\doibase 10.1103/PhysRevD.103.014513} {\bibfield  {journal} {\bibinfo  {journal} {Phys. Rev. D}\ }\textbf {\bibinfo {volume} {103}},\ \bibinfo {pages} {014513} (\bibinfo {year} {2021}{\natexlab{b}})},\ \Eprint {http://arxiv.org/abs/2009.14033} {arXiv:2009.14033 [hep-lat]} \BibitemShut {NoStop}%
\bibitem [{\citenamefont {Bazavov}\ \emph {et~al.}(2009)\citenamefont {Bazavov}, \citenamefont {Bhattacharya}, \citenamefont {Cheng}, \citenamefont {Christ}, \citenamefont {DeTar} \emph {et~al.}}]{Bazavov:2009zn}%
  \BibitemOpen
  \bibfield  {author} {\bibinfo {author} {\bibfnamefont {A.}~\bibnamefont {Bazavov}}, \bibinfo {author} {\bibfnamefont {T.}~\bibnamefont {Bhattacharya}}, \bibinfo {author} {\bibfnamefont {M.}~\bibnamefont {Cheng}}, \bibinfo {author} {\bibfnamefont {N.}~\bibnamefont {Christ}}, \bibinfo {author} {\bibfnamefont {C.}~\bibnamefont {DeTar}},  \emph {et~al.},\ }\href {\doibase 10.1103/PhysRevD.80.014504} {\bibfield  {journal} {\bibinfo  {journal} {Phys.Rev.}\ }\textbf {\bibinfo {volume} {D80}},\ \bibinfo {pages} {014504} (\bibinfo {year} {2009})},\ \Eprint {http://arxiv.org/abs/0903.4379} {arXiv:0903.4379 [hep-lat]} \BibitemShut {NoStop}%
\bibitem [{\citenamefont {Bazavov}\ and\ \citenamefont {Petreczky}(2013)}]{Bazavov:2013yv}%
  \BibitemOpen
  \bibfield  {author} {\bibinfo {author} {\bibfnamefont {A.}~\bibnamefont {Bazavov}}\ and\ \bibinfo {author} {\bibfnamefont {P.}~\bibnamefont {Petreczky}},\ }\href {\doibase 10.1103/PhysRevD.87.094505} {\bibfield  {journal} {\bibinfo  {journal} {Phys.Rev.}\ }\textbf {\bibinfo {volume} {D87}},\ \bibinfo {pages} {094505} (\bibinfo {year} {2013})},\ \Eprint {http://arxiv.org/abs/1301.3943} {arXiv:1301.3943 [hep-lat]} \BibitemShut {NoStop}%
\bibitem [{\citenamefont {Bors{\'a}nyi}\ \emph {et~al.}(2015)\citenamefont {Bors{\'a}nyi}, \citenamefont {Fodor}, \citenamefont {Katz}, \citenamefont {P{\'a}sztor}, \citenamefont {Szab{\'o}} \emph {et~al.}}]{Borsanyi:2015yka}%
  \BibitemOpen
  \bibfield  {author} {\bibinfo {author} {\bibfnamefont {S.}~\bibnamefont {Bors{\'a}nyi}}, \bibinfo {author} {\bibfnamefont {Z.}~\bibnamefont {Fodor}}, \bibinfo {author} {\bibfnamefont {S.~D.}\ \bibnamefont {Katz}}, \bibinfo {author} {\bibfnamefont {A.}~\bibnamefont {P{\'a}sztor}}, \bibinfo {author} {\bibfnamefont {K.~K.}\ \bibnamefont {Szab{\'o}}},  \emph {et~al.},\ }\href {\doibase 10.1007/JHEP04(2015)138} {\bibfield  {journal} {\bibinfo  {journal} {JHEP}\ }\textbf {\bibinfo {volume} {1504}},\ \bibinfo {pages} {138} (\bibinfo {year} {2015})},\ \Eprint {http://arxiv.org/abs/1501.02173} {arXiv:1501.02173 [hep-lat]} \BibitemShut {NoStop}%
\bibitem [{\citenamefont {Clarke}\ \emph {et~al.}(2021)\citenamefont {Clarke}, \citenamefont {Kaczmarek}, \citenamefont {Karsch}, \citenamefont {Lahiri},\ and\ \citenamefont {Sarkar}}]{Clarke:2020htu}%
  \BibitemOpen
  \bibfield  {author} {\bibinfo {author} {\bibfnamefont {D.~A.}\ \bibnamefont {Clarke}}, \bibinfo {author} {\bibfnamefont {O.}~\bibnamefont {Kaczmarek}}, \bibinfo {author} {\bibfnamefont {F.}~\bibnamefont {Karsch}}, \bibinfo {author} {\bibfnamefont {A.}~\bibnamefont {Lahiri}}, \ and\ \bibinfo {author} {\bibfnamefont {M.}~\bibnamefont {Sarkar}},\ }\href {\doibase 10.1103/PhysRevD.103.L011501} {\bibfield  {journal} {\bibinfo  {journal} {Phys. Rev. D}\ }\textbf {\bibinfo {volume} {103}},\ \bibinfo {pages} {L011501} (\bibinfo {year} {2021})},\ \Eprint {http://arxiv.org/abs/2008.11678} {arXiv:2008.11678 [hep-lat]} \BibitemShut {NoStop}%
\bibitem [{\citenamefont {Bazavov}\ \emph {et~al.}(2016)\citenamefont {Bazavov}, \citenamefont {Brambilla}, \citenamefont {Ding}, \citenamefont {Petreczky}, \citenamefont {Schadler}, \citenamefont {Vairo},\ and\ \citenamefont {Weber}}]{Bazavov:2016uvm}%
  \BibitemOpen
  \bibfield  {author} {\bibinfo {author} {\bibfnamefont {A.}~\bibnamefont {Bazavov}}, \bibinfo {author} {\bibfnamefont {N.}~\bibnamefont {Brambilla}}, \bibinfo {author} {\bibfnamefont {H.~T.}\ \bibnamefont {Ding}}, \bibinfo {author} {\bibfnamefont {P.}~\bibnamefont {Petreczky}}, \bibinfo {author} {\bibfnamefont {H.~P.}\ \bibnamefont {Schadler}}, \bibinfo {author} {\bibfnamefont {A.}~\bibnamefont {Vairo}}, \ and\ \bibinfo {author} {\bibfnamefont {J.~H.}\ \bibnamefont {Weber}},\ }\href {\doibase 10.1103/PhysRevD.93.114502} {\bibfield  {journal} {\bibinfo  {journal} {Phys. Rev.}\ }\textbf {\bibinfo {volume} {D93}},\ \bibinfo {pages} {114502} (\bibinfo {year} {2016})},\ \Eprint {http://arxiv.org/abs/1603.06637} {arXiv:1603.06637 [hep-lat]} \BibitemShut {NoStop}%
\bibitem [{\citenamefont {Kov\'acs}\ \emph {et~al.}(2016)\citenamefont {Kov\'acs}, \citenamefont {Sz\'ep},\ and\ \citenamefont {Wolf}}]{Kovacs:2016juc}%
  \BibitemOpen
  \bibfield  {author} {\bibinfo {author} {\bibfnamefont {P.}~\bibnamefont {Kov\'acs}}, \bibinfo {author} {\bibfnamefont {Z.}~\bibnamefont {Sz\'ep}}, \ and\ \bibinfo {author} {\bibfnamefont {G.}~\bibnamefont {Wolf}},\ }\href {\doibase 10.1103/PhysRevD.93.114014} {\bibfield  {journal} {\bibinfo  {journal} {Phys. Rev. D}\ }\textbf {\bibinfo {volume} {93}},\ \bibinfo {pages} {114014} (\bibinfo {year} {2016})},\ \Eprint {http://arxiv.org/abs/1601.05291} {arXiv:1601.05291 [hep-ph]} \BibitemShut {NoStop}%
\bibitem [{\citenamefont {Gao}\ and\ \citenamefont {Pawlowski}(2020)}]{Gao:2020qsj}%
  \BibitemOpen
  \bibfield  {author} {\bibinfo {author} {\bibfnamefont {F.}~\bibnamefont {Gao}}\ and\ \bibinfo {author} {\bibfnamefont {J.~M.}\ \bibnamefont {Pawlowski}},\ }\href {\doibase 10.1103/PhysRevD.102.034027} {\bibfield  {journal} {\bibinfo  {journal} {Phys. Rev. D}\ }\textbf {\bibinfo {volume} {102}},\ \bibinfo {pages} {034027} (\bibinfo {year} {2020})},\ \Eprint {http://arxiv.org/abs/2002.07500} {arXiv:2002.07500 [hep-ph]} \BibitemShut {NoStop}%
\bibitem [{\citenamefont {Fu}\ \emph {et~al.}(2021)\citenamefont {Fu}, \citenamefont {Luo}, \citenamefont {Pawlowski}, \citenamefont {Rennecke}, \citenamefont {Wen},\ and\ \citenamefont {Yin}}]{Fu:2021oaw}%
  \BibitemOpen
  \bibfield  {author} {\bibinfo {author} {\bibfnamefont {W.-j.}\ \bibnamefont {Fu}}, \bibinfo {author} {\bibfnamefont {X.}~\bibnamefont {Luo}}, \bibinfo {author} {\bibfnamefont {J.~M.}\ \bibnamefont {Pawlowski}}, \bibinfo {author} {\bibfnamefont {F.}~\bibnamefont {Rennecke}}, \bibinfo {author} {\bibfnamefont {R.}~\bibnamefont {Wen}}, \ and\ \bibinfo {author} {\bibfnamefont {S.}~\bibnamefont {Yin}},\ }\href {\doibase 10.1103/PhysRevD.104.094047} {\bibfield  {journal} {\bibinfo  {journal} {Phys. Rev. D}\ }\textbf {\bibinfo {volume} {104}},\ \bibinfo {pages} {094047} (\bibinfo {year} {2021})},\ \Eprint {http://arxiv.org/abs/2101.06035} {arXiv:2101.06035 [hep-ph]} \BibitemShut {NoStop}%
\bibitem [{\citenamefont {Isserstedt}\ \emph {et~al.}(2019)\citenamefont {Isserstedt}, \citenamefont {Buballa}, \citenamefont {Fischer},\ and\ \citenamefont {Gunkel}}]{Isserstedt:2019pgx}%
  \BibitemOpen
  \bibfield  {author} {\bibinfo {author} {\bibfnamefont {P.}~\bibnamefont {Isserstedt}}, \bibinfo {author} {\bibfnamefont {M.}~\bibnamefont {Buballa}}, \bibinfo {author} {\bibfnamefont {C.~S.}\ \bibnamefont {Fischer}}, \ and\ \bibinfo {author} {\bibfnamefont {P.~J.}\ \bibnamefont {Gunkel}},\ }\href {\doibase 10.1103/PhysRevD.100.074011} {\bibfield  {journal} {\bibinfo  {journal} {Phys. Rev.}\ }\textbf {\bibinfo {volume} {D100}},\ \bibinfo {pages} {074011} (\bibinfo {year} {2019})},\ \Eprint {http://arxiv.org/abs/1906.11644} {arXiv:1906.11644 [hep-ph]} \BibitemShut {NoStop}%
\bibitem [{\citenamefont {Critelli}\ \emph {et~al.}(2017)\citenamefont {Critelli}, \citenamefont {Noronha}, \citenamefont {Noronha-Hostler}, \citenamefont {Portillo}, \citenamefont {Ratti},\ and\ \citenamefont {Rougemont}}]{Critelli:2017oub}%
  \BibitemOpen
  \bibfield  {author} {\bibinfo {author} {\bibfnamefont {R.}~\bibnamefont {Critelli}}, \bibinfo {author} {\bibfnamefont {J.}~\bibnamefont {Noronha}}, \bibinfo {author} {\bibfnamefont {J.}~\bibnamefont {Noronha-Hostler}}, \bibinfo {author} {\bibfnamefont {I.}~\bibnamefont {Portillo}}, \bibinfo {author} {\bibfnamefont {C.}~\bibnamefont {Ratti}}, \ and\ \bibinfo {author} {\bibfnamefont {R.}~\bibnamefont {Rougemont}},\ }\href {\doibase 10.1103/PhysRevD.96.096026} {\bibfield  {journal} {\bibinfo  {journal} {Phys. Rev.}\ }\textbf {\bibinfo {volume} {D96}},\ \bibinfo {pages} {096026} (\bibinfo {year} {2017})},\ \Eprint {http://arxiv.org/abs/1706.00455} {arXiv:1706.00455 [nucl-th]} \BibitemShut {NoStop}%
\bibitem [{\citenamefont {Hippert}\ \emph {et~al.}(2023)\citenamefont {Hippert}, \citenamefont {Grefa}, \citenamefont {Manning}, \citenamefont {Noronha}, \citenamefont {Noronha-Hostler}, \citenamefont {Portillo~Vazquez}, \citenamefont {Ratti}, \citenamefont {Rougemont},\ and\ \citenamefont {Trujillo}}]{Hippert:2023bel}%
  \BibitemOpen
  \bibfield  {author} {\bibinfo {author} {\bibfnamefont {M.}~\bibnamefont {Hippert}}, \bibinfo {author} {\bibfnamefont {J.}~\bibnamefont {Grefa}}, \bibinfo {author} {\bibfnamefont {T.~A.}\ \bibnamefont {Manning}}, \bibinfo {author} {\bibfnamefont {J.}~\bibnamefont {Noronha}}, \bibinfo {author} {\bibfnamefont {J.}~\bibnamefont {Noronha-Hostler}}, \bibinfo {author} {\bibfnamefont {I.}~\bibnamefont {Portillo~Vazquez}}, \bibinfo {author} {\bibfnamefont {C.}~\bibnamefont {Ratti}}, \bibinfo {author} {\bibfnamefont {R.}~\bibnamefont {Rougemont}}, \ and\ \bibinfo {author} {\bibfnamefont {M.}~\bibnamefont {Trujillo}},\ }\href@noop {} {\  (\bibinfo {year} {2023})},\ \Eprint {http://arxiv.org/abs/2309.00579} {arXiv:2309.00579 [nucl-th]} \BibitemShut {NoStop}%
\bibitem [{\citenamefont {Adam}\ \emph {et~al.}(2021)\citenamefont {Adam} \emph {et~al.}}]{STAR:2020tga}%
  \BibitemOpen
  \bibfield  {author} {\bibinfo {author} {\bibfnamefont {J.}~\bibnamefont {Adam}} \emph {et~al.} (\bibinfo {collaboration} {STAR}),\ }\href {\doibase 10.1103/PhysRevLett.126.092301} {\bibfield  {journal} {\bibinfo  {journal} {Phys. Rev. Lett.}\ }\textbf {\bibinfo {volume} {126}},\ \bibinfo {pages} {092301} (\bibinfo {year} {2021})},\ \Eprint {http://arxiv.org/abs/2001.02852} {arXiv:2001.02852 [nucl-ex]} \BibitemShut {NoStop}%
\bibitem [{\citenamefont {Bonati}\ \emph {et~al.}(2018)\citenamefont {Bonati}, \citenamefont {D'Elia}, \citenamefont {Negro}, \citenamefont {Sanfilippo},\ and\ \citenamefont {Zambello}}]{Bonati:2018nut}%
  \BibitemOpen
  \bibfield  {author} {\bibinfo {author} {\bibfnamefont {C.}~\bibnamefont {Bonati}}, \bibinfo {author} {\bibfnamefont {M.}~\bibnamefont {D'Elia}}, \bibinfo {author} {\bibfnamefont {F.}~\bibnamefont {Negro}}, \bibinfo {author} {\bibfnamefont {F.}~\bibnamefont {Sanfilippo}}, \ and\ \bibinfo {author} {\bibfnamefont {K.}~\bibnamefont {Zambello}},\ }\href {\doibase 10.1103/PhysRevD.98.054510} {\bibfield  {journal} {\bibinfo  {journal} {Phys. Rev.}\ }\textbf {\bibinfo {volume} {D98}},\ \bibinfo {pages} {054510} (\bibinfo {year} {2018})},\ \Eprint {http://arxiv.org/abs/1805.02960} {arXiv:1805.02960 [hep-lat]} \BibitemShut {NoStop}%
\bibitem [{\citenamefont {Bazavov}\ \emph {et~al.}(2019{\natexlab{a}})\citenamefont {Bazavov} \emph {et~al.}}]{HotQCD:2018pds}%
  \BibitemOpen
  \bibfield  {author} {\bibinfo {author} {\bibfnamefont {A.}~\bibnamefont {Bazavov}} \emph {et~al.} (\bibinfo {collaboration} {HotQCD}),\ }\href {\doibase 10.1016/j.physletb.2019.05.013} {\bibfield  {journal} {\bibinfo  {journal} {Phys. Lett. B}\ }\textbf {\bibinfo {volume} {795}},\ \bibinfo {pages} {15} (\bibinfo {year} {2019}{\natexlab{a}})},\ \Eprint {http://arxiv.org/abs/1812.08235} {arXiv:1812.08235 [hep-lat]} \BibitemShut {NoStop}%
\bibitem [{\citenamefont {P\'asztor}\ \emph {et~al.}(2021)\citenamefont {P\'asztor}, \citenamefont {Sz\'ep},\ and\ \citenamefont {Mark\'o}}]{Pasztor:2020dur}%
  \BibitemOpen
  \bibfield  {author} {\bibinfo {author} {\bibfnamefont {A.}~\bibnamefont {P\'asztor}}, \bibinfo {author} {\bibfnamefont {Z.}~\bibnamefont {Sz\'ep}}, \ and\ \bibinfo {author} {\bibfnamefont {G.}~\bibnamefont {Mark\'o}},\ }\href {\doibase 10.1103/PhysRevD.103.034511} {\bibfield  {journal} {\bibinfo  {journal} {Phys. Rev. D}\ }\textbf {\bibinfo {volume} {103}},\ \bibinfo {pages} {034511} (\bibinfo {year} {2021})},\ \Eprint {http://arxiv.org/abs/2010.00394} {arXiv:2010.00394 [hep-lat]} \BibitemShut {NoStop}%
\bibitem [{\citenamefont {D'Elia}\ \emph {et~al.}(2019)\citenamefont {D'Elia}, \citenamefont {Negro}, \citenamefont {Rucci},\ and\ \citenamefont {Sanfilippo}}]{DElia:2019iis}%
  \BibitemOpen
  \bibfield  {author} {\bibinfo {author} {\bibfnamefont {M.}~\bibnamefont {D'Elia}}, \bibinfo {author} {\bibfnamefont {F.}~\bibnamefont {Negro}}, \bibinfo {author} {\bibfnamefont {A.}~\bibnamefont {Rucci}}, \ and\ \bibinfo {author} {\bibfnamefont {F.}~\bibnamefont {Sanfilippo}},\ }\href {\doibase 10.1103/PhysRevD.100.054504} {\bibfield  {journal} {\bibinfo  {journal} {Phys. Rev. D}\ }\textbf {\bibinfo {volume} {100}},\ \bibinfo {pages} {054504} (\bibinfo {year} {2019})},\ \Eprint {http://arxiv.org/abs/1907.09461} {arXiv:1907.09461 [hep-lat]} \BibitemShut {NoStop}%
\bibitem [{\citenamefont {Borsanyi}\ \emph {et~al.}(2024{\natexlab{a}})\citenamefont {Borsanyi}, \citenamefont {Fodor}, \citenamefont {Guenther}, \citenamefont {Kara}, \citenamefont {Parotto}, \citenamefont {Pasztor}, \citenamefont {Pirelli},\ and\ \citenamefont {Wong}}]{Borsanyi:2024dro}%
  \BibitemOpen
  \bibfield  {author} {\bibinfo {author} {\bibfnamefont {S.}~\bibnamefont {Borsanyi}}, \bibinfo {author} {\bibfnamefont {Z.}~\bibnamefont {Fodor}}, \bibinfo {author} {\bibfnamefont {J.~N.}\ \bibnamefont {Guenther}}, \bibinfo {author} {\bibfnamefont {R.}~\bibnamefont {Kara}}, \bibinfo {author} {\bibfnamefont {P.}~\bibnamefont {Parotto}}, \bibinfo {author} {\bibfnamefont {A.}~\bibnamefont {Pasztor}}, \bibinfo {author} {\bibfnamefont {L.}~\bibnamefont {Pirelli}}, \ and\ \bibinfo {author} {\bibfnamefont {C.~H.}\ \bibnamefont {Wong}},\ }\href@noop {} {\  (\bibinfo {year} {2024}{\natexlab{a}})},\ \Eprint {http://arxiv.org/abs/2405.12320} {arXiv:2405.12320 [hep-lat]} \BibitemShut {NoStop}%
\bibitem [{\citenamefont {Borsanyi}\ \emph {et~al.}(2024{\natexlab{b}})\citenamefont {Borsanyi}, \citenamefont {Fodor}, \citenamefont {Guenther}, \citenamefont {Katz}, \citenamefont {Parotto}, \citenamefont {Pasztor}, \citenamefont {Pesznyak}, \citenamefont {Szabo},\ and\ \citenamefont {Wong}}]{Borsanyi:2023wno}%
  \BibitemOpen
  \bibfield  {author} {\bibinfo {author} {\bibfnamefont {S.}~\bibnamefont {Borsanyi}}, \bibinfo {author} {\bibfnamefont {Z.}~\bibnamefont {Fodor}}, \bibinfo {author} {\bibfnamefont {J.~N.}\ \bibnamefont {Guenther}}, \bibinfo {author} {\bibfnamefont {S.~D.}\ \bibnamefont {Katz}}, \bibinfo {author} {\bibfnamefont {P.}~\bibnamefont {Parotto}}, \bibinfo {author} {\bibfnamefont {A.}~\bibnamefont {Pasztor}}, \bibinfo {author} {\bibfnamefont {D.}~\bibnamefont {Pesznyak}}, \bibinfo {author} {\bibfnamefont {K.~K.}\ \bibnamefont {Szabo}}, \ and\ \bibinfo {author} {\bibfnamefont {C.~H.}\ \bibnamefont {Wong}},\ }\href {\doibase 10.1103/PhysRevD.110.L011501} {\bibfield  {journal} {\bibinfo  {journal} {Phys. Rev. D}\ }\textbf {\bibinfo {volume} {110}},\ \bibinfo {pages} {L011501} (\bibinfo {year} {2024}{\natexlab{b}})},\ \Eprint {http://arxiv.org/abs/2312.07528} {arXiv:2312.07528 [hep-lat]} \BibitemShut {NoStop}%
\bibitem [{\citenamefont {Allton}\ \emph {et~al.}(2005)\citenamefont {Allton}, \citenamefont {Doring}, \citenamefont {Ejiri}, \citenamefont {Hands}, \citenamefont {Kaczmarek} \emph {et~al.}}]{Allton:2005gk}%
  \BibitemOpen
  \bibfield  {author} {\bibinfo {author} {\bibfnamefont {C.}~\bibnamefont {Allton}}, \bibinfo {author} {\bibfnamefont {M.}~\bibnamefont {Doring}}, \bibinfo {author} {\bibfnamefont {S.}~\bibnamefont {Ejiri}}, \bibinfo {author} {\bibfnamefont {S.}~\bibnamefont {Hands}}, \bibinfo {author} {\bibfnamefont {O.}~\bibnamefont {Kaczmarek}},  \emph {et~al.},\ }\href {\doibase 10.1103/PhysRevD.71.054508} {\bibfield  {journal} {\bibinfo  {journal} {Phys.Rev.}\ }\textbf {\bibinfo {volume} {D71}},\ \bibinfo {pages} {054508} (\bibinfo {year} {2005})},\ \Eprint {http://arxiv.org/abs/hep-lat/0501030} {arXiv:hep-lat/0501030 [hep-lat]} \BibitemShut {NoStop}%
\bibitem [{\citenamefont {Bazavov}\ \emph {et~al.}(2017)\citenamefont {Bazavov} \emph {et~al.}}]{Bazavov:2017dus}%
  \BibitemOpen
  \bibfield  {author} {\bibinfo {author} {\bibfnamefont {A.}~\bibnamefont {Bazavov}} \emph {et~al.},\ }\href {\doibase 10.1103/PhysRevD.95.054504} {\bibfield  {journal} {\bibinfo  {journal} {Phys. Rev.}\ }\textbf {\bibinfo {volume} {D95}},\ \bibinfo {pages} {054504} (\bibinfo {year} {2017})},\ \Eprint {http://arxiv.org/abs/1701.04325} {arXiv:1701.04325 [hep-lat]} \BibitemShut {NoStop}%
\bibitem [{\citenamefont {Borsanyi}\ \emph {et~al.}(2023)\citenamefont {Borsanyi}, \citenamefont {Fodor}, \citenamefont {Giordano}, \citenamefont {Guenther}, \citenamefont {Katz}, \citenamefont {Pasztor},\ and\ \citenamefont {Wong}}]{Borsanyi:2022soo}%
  \BibitemOpen
  \bibfield  {author} {\bibinfo {author} {\bibfnamefont {S.}~\bibnamefont {Borsanyi}}, \bibinfo {author} {\bibfnamefont {Z.}~\bibnamefont {Fodor}}, \bibinfo {author} {\bibfnamefont {M.}~\bibnamefont {Giordano}}, \bibinfo {author} {\bibfnamefont {J.~N.}\ \bibnamefont {Guenther}}, \bibinfo {author} {\bibfnamefont {S.~D.}\ \bibnamefont {Katz}}, \bibinfo {author} {\bibfnamefont {A.}~\bibnamefont {Pasztor}}, \ and\ \bibinfo {author} {\bibfnamefont {C.~H.}\ \bibnamefont {Wong}},\ }\href {\doibase 10.1103/PhysRevD.107.L091503} {\bibfield  {journal} {\bibinfo  {journal} {Phys. Rev. D}\ }\textbf {\bibinfo {volume} {107}},\ \bibinfo {pages} {L091503} (\bibinfo {year} {2023})},\ \Eprint {http://arxiv.org/abs/2208.05398} {arXiv:2208.05398 [hep-lat]} \BibitemShut {NoStop}%
\bibitem [{\citenamefont {Borsanyi}\ \emph {et~al.}(2024{\natexlab{c}})\citenamefont {Borsanyi}, \citenamefont {Fodor}, \citenamefont {Giordano}, \citenamefont {Guenther}, \citenamefont {Katz}, \citenamefont {Pasztor},\ and\ \citenamefont {Wong}}]{Borsanyi:2023tdp}%
  \BibitemOpen
  \bibfield  {author} {\bibinfo {author} {\bibfnamefont {S.}~\bibnamefont {Borsanyi}}, \bibinfo {author} {\bibfnamefont {Z.}~\bibnamefont {Fodor}}, \bibinfo {author} {\bibfnamefont {M.}~\bibnamefont {Giordano}}, \bibinfo {author} {\bibfnamefont {J.~N.}\ \bibnamefont {Guenther}}, \bibinfo {author} {\bibfnamefont {S.~D.}\ \bibnamefont {Katz}}, \bibinfo {author} {\bibfnamefont {A.}~\bibnamefont {Pasztor}}, \ and\ \bibinfo {author} {\bibfnamefont {C.~H.}\ \bibnamefont {Wong}},\ }\href {\doibase 10.1103/PhysRevD.109.054509} {\bibfield  {journal} {\bibinfo  {journal} {Phys. Rev. D}\ }\textbf {\bibinfo {volume} {109}},\ \bibinfo {pages} {054509} (\bibinfo {year} {2024}{\natexlab{c}})},\ \Eprint {http://arxiv.org/abs/2308.06105} {arXiv:2308.06105 [hep-lat]} \BibitemShut {NoStop}%
\bibitem [{\citenamefont {Hasenfratz}\ and\ \citenamefont {Toussaint}(1992)}]{Hasenfratz:1991ax}%
  \BibitemOpen
  \bibfield  {author} {\bibinfo {author} {\bibfnamefont {A.}~\bibnamefont {Hasenfratz}}\ and\ \bibinfo {author} {\bibfnamefont {D.}~\bibnamefont {Toussaint}},\ }\href {\doibase 10.1016/0550-3213(92)90247-9} {\bibfield  {journal} {\bibinfo  {journal} {Nucl. Phys.}\ }\textbf {\bibinfo {volume} {B371}},\ \bibinfo {pages} {539} (\bibinfo {year} {1992})}\BibitemShut {NoStop}%
\bibitem [{\citenamefont {Hasenfratz}\ and\ \citenamefont {Karsch}(1983)}]{Hasenfratz:1983ba}%
  \BibitemOpen
  \bibfield  {author} {\bibinfo {author} {\bibfnamefont {P.}~\bibnamefont {Hasenfratz}}\ and\ \bibinfo {author} {\bibfnamefont {F.}~\bibnamefont {Karsch}},\ }\href {\doibase 10.1016/0370-2693(83)91290-X} {\bibfield  {journal} {\bibinfo  {journal} {Phys.Lett.}\ }\textbf {\bibinfo {volume} {B125}},\ \bibinfo {pages} {308} (\bibinfo {year} {1983})}\BibitemShut {NoStop}%
\bibitem [{\citenamefont {Allton}\ \emph {et~al.}(2002)\citenamefont {Allton}, \citenamefont {Ejiri}, \citenamefont {Hands}, \citenamefont {Kaczmarek}, \citenamefont {Karsch} \emph {et~al.}}]{Allton:2002zi}%
  \BibitemOpen
  \bibfield  {author} {\bibinfo {author} {\bibfnamefont {C.}~\bibnamefont {Allton}}, \bibinfo {author} {\bibfnamefont {S.}~\bibnamefont {Ejiri}}, \bibinfo {author} {\bibfnamefont {S.}~\bibnamefont {Hands}}, \bibinfo {author} {\bibfnamefont {O.}~\bibnamefont {Kaczmarek}}, \bibinfo {author} {\bibfnamefont {F.}~\bibnamefont {Karsch}},  \emph {et~al.},\ }\href {\doibase 10.1103/PhysRevD.66.074507} {\bibfield  {journal} {\bibinfo  {journal} {Phys.Rev.}\ }\textbf {\bibinfo {volume} {D66}},\ \bibinfo {pages} {074507} (\bibinfo {year} {2002})},\ \Eprint {http://arxiv.org/abs/hep-lat/0204010} {arXiv:hep-lat/0204010 [hep-lat]} \BibitemShut {NoStop}%
\bibitem [{\citenamefont {Tomov}\ \emph {et~al.}(2010)\citenamefont {Tomov}, \citenamefont {Dongarra},\ and\ \citenamefont {Baboulin}}]{tdb10}%
  \BibitemOpen
  \bibfield  {author} {\bibinfo {author} {\bibfnamefont {S.}~\bibnamefont {Tomov}}, \bibinfo {author} {\bibfnamefont {J.}~\bibnamefont {Dongarra}}, \ and\ \bibinfo {author} {\bibfnamefont {M.}~\bibnamefont {Baboulin}},\ }\bibfield  {booktitle} {\emph {\bibinfo {booktitle} {Parallel Matrix Algorithms and Applications}},\ }\href {\doibase 10.1016/j.parco.2009.12.005} {\bibfield  {journal} {\bibinfo  {journal} {Parallel Computing}\ }\textbf {\bibinfo {volume} {36}},\ \bibinfo {pages} {232} (\bibinfo {year} {2010})}\BibitemShut {NoStop}%
\bibitem [{\citenamefont {Bosma}\ \emph {et~al.}(1997)\citenamefont {Bosma}, \citenamefont {Cannon},\ and\ \citenamefont {Playoust}}]{MR1484478}%
  \BibitemOpen
  \bibfield  {author} {\bibinfo {author} {\bibfnamefont {W.}~\bibnamefont {Bosma}}, \bibinfo {author} {\bibfnamefont {J.}~\bibnamefont {Cannon}}, \ and\ \bibinfo {author} {\bibfnamefont {C.}~\bibnamefont {Playoust}},\ }\href {\doibase 10.1006/jsco.1996.0125} {\bibfield  {journal} {\bibinfo  {journal} {J. Symbolic Comput.}\ }\textbf {\bibinfo {volume} {24}},\ \bibinfo {pages} {235} (\bibinfo {year} {1997})},\ \bibinfo {note} {computational algebra and number theory (London, 1993)}\BibitemShut {NoStop}%
\bibitem [{\citenamefont {Brown}\ \emph {et~al.}(2020)\citenamefont {Brown}, \citenamefont {Abdelfattah}, \citenamefont {Tomov},\ and\ \citenamefont {Dongarra}}]{BrownATD20}%
  \BibitemOpen
  \bibfield  {author} {\bibinfo {author} {\bibfnamefont {C.}~\bibnamefont {Brown}}, \bibinfo {author} {\bibfnamefont {A.}~\bibnamefont {Abdelfattah}}, \bibinfo {author} {\bibfnamefont {S.}~\bibnamefont {Tomov}}, \ and\ \bibinfo {author} {\bibfnamefont {J.~J.}\ \bibnamefont {Dongarra}},\ }in\ \href {\doibase 10.1109/HPEC43674.2020.9286214} {\emph {\bibinfo {booktitle} {2020 {IEEE} High Performance Extreme Computing Conference, {HPEC} 2020, Waltham, MA, USA, September 22-24, 2020}}}\ (\bibinfo  {publisher} {{IEEE}},\ \bibinfo {year} {2020})\ pp.\ \bibinfo {pages} {1--7}\BibitemShut {NoStop}%
\bibitem [{\citenamefont {Gupta}\ \emph {et~al.}(2008)\citenamefont {Gupta}, \citenamefont {Huebner},\ and\ \citenamefont {Kaczmarek}}]{Gupta:2007ax}%
  \BibitemOpen
  \bibfield  {author} {\bibinfo {author} {\bibfnamefont {S.}~\bibnamefont {Gupta}}, \bibinfo {author} {\bibfnamefont {K.}~\bibnamefont {Huebner}}, \ and\ \bibinfo {author} {\bibfnamefont {O.}~\bibnamefont {Kaczmarek}},\ }\href {\doibase 10.1103/PhysRevD.77.034503} {\bibfield  {journal} {\bibinfo  {journal} {Phys.Rev.}\ }\textbf {\bibinfo {volume} {D77}},\ \bibinfo {pages} {034503} (\bibinfo {year} {2008})},\ \Eprint {http://arxiv.org/abs/0711.2251} {arXiv:0711.2251 [hep-lat]} \BibitemShut {NoStop}%
\bibitem [{\citenamefont {Cheng}\ \emph {et~al.}(2008)\citenamefont {Cheng}, \citenamefont {Christ}, \citenamefont {Datta}, \citenamefont {van~der Heide}, \citenamefont {Jung} \emph {et~al.}}]{Cheng:2007jq}%
  \BibitemOpen
  \bibfield  {author} {\bibinfo {author} {\bibfnamefont {M.}~\bibnamefont {Cheng}}, \bibinfo {author} {\bibfnamefont {N.}~\bibnamefont {Christ}}, \bibinfo {author} {\bibfnamefont {S.}~\bibnamefont {Datta}}, \bibinfo {author} {\bibfnamefont {J.}~\bibnamefont {van~der Heide}}, \bibinfo {author} {\bibfnamefont {C.}~\bibnamefont {Jung}},  \emph {et~al.},\ }\href {\doibase 10.1103/PhysRevD.77.014511} {\bibfield  {journal} {\bibinfo  {journal} {Phys.Rev.}\ }\textbf {\bibinfo {volume} {D77}},\ \bibinfo {pages} {014511} (\bibinfo {year} {2008})},\ \Eprint {http://arxiv.org/abs/0710.0354} {arXiv:0710.0354 [hep-lat]} \BibitemShut {NoStop}%
\bibitem [{\citenamefont {Kaczmarek}\ \emph {et~al.}(2002)\citenamefont {Kaczmarek}, \citenamefont {Karsch}, \citenamefont {Petreczky},\ and\ \citenamefont {Zantow}}]{Kaczmarek:2002mc}%
  \BibitemOpen
  \bibfield  {author} {\bibinfo {author} {\bibfnamefont {O.}~\bibnamefont {Kaczmarek}}, \bibinfo {author} {\bibfnamefont {F.}~\bibnamefont {Karsch}}, \bibinfo {author} {\bibfnamefont {P.}~\bibnamefont {Petreczky}}, \ and\ \bibinfo {author} {\bibfnamefont {F.}~\bibnamefont {Zantow}},\ }\href {\doibase 10.1016/S0370-2693(02)02415-2} {\bibfield  {journal} {\bibinfo  {journal} {Phys.Lett.}\ }\textbf {\bibinfo {volume} {B543}},\ \bibinfo {pages} {41} (\bibinfo {year} {2002})},\ \Eprint {http://arxiv.org/abs/hep-lat/0207002} {arXiv:hep-lat/0207002 [hep-lat]} \BibitemShut {NoStop}%
\bibitem [{\citenamefont {Petreczky}\ and\ \citenamefont {Schadler}(2015)}]{Petreczky:2015yta}%
  \BibitemOpen
  \bibfield  {author} {\bibinfo {author} {\bibfnamefont {P.}~\bibnamefont {Petreczky}}\ and\ \bibinfo {author} {\bibfnamefont {H.~P.}\ \bibnamefont {Schadler}},\ }\href {\doibase 10.1103/PhysRevD.92.094517} {\bibfield  {journal} {\bibinfo  {journal} {Phys. Rev.}\ }\textbf {\bibinfo {volume} {D92}},\ \bibinfo {pages} {094517} (\bibinfo {year} {2015})},\ \Eprint {http://arxiv.org/abs/1509.07874} {arXiv:1509.07874 [hep-lat]} \BibitemShut {NoStop}%
\bibitem [{\citenamefont {Datta}\ \emph {et~al.}(2016)\citenamefont {Datta}, \citenamefont {Gupta},\ and\ \citenamefont {Lytle}}]{Datta:2015bzm}%
  \BibitemOpen
  \bibfield  {author} {\bibinfo {author} {\bibfnamefont {S.}~\bibnamefont {Datta}}, \bibinfo {author} {\bibfnamefont {S.}~\bibnamefont {Gupta}}, \ and\ \bibinfo {author} {\bibfnamefont {A.}~\bibnamefont {Lytle}},\ }\href {\doibase 10.1103/PhysRevD.94.094502} {\bibfield  {journal} {\bibinfo  {journal} {Phys. Rev. D}\ }\textbf {\bibinfo {volume} {94}},\ \bibinfo {pages} {094502} (\bibinfo {year} {2016})},\ \Eprint {http://arxiv.org/abs/1512.04892} {arXiv:1512.04892 [hep-lat]} \BibitemShut {NoStop}%
\bibitem [{\citenamefont {Bonati}\ \emph {et~al.}(2016)\citenamefont {Bonati}, \citenamefont {D'Elia}, \citenamefont {Mariti}, \citenamefont {Mesiti}, \citenamefont {Negro},\ and\ \citenamefont {Sanfilippo}}]{Bonati:2016pwz}%
  \BibitemOpen
  \bibfield  {author} {\bibinfo {author} {\bibfnamefont {C.}~\bibnamefont {Bonati}}, \bibinfo {author} {\bibfnamefont {M.}~\bibnamefont {D'Elia}}, \bibinfo {author} {\bibfnamefont {M.}~\bibnamefont {Mariti}}, \bibinfo {author} {\bibfnamefont {M.}~\bibnamefont {Mesiti}}, \bibinfo {author} {\bibfnamefont {F.}~\bibnamefont {Negro}}, \ and\ \bibinfo {author} {\bibfnamefont {F.}~\bibnamefont {Sanfilippo}},\ }\href {\doibase 10.1103/PhysRevD.93.074504} {\bibfield  {journal} {\bibinfo  {journal} {Phys. Rev.}\ }\textbf {\bibinfo {volume} {D93}},\ \bibinfo {pages} {074504} (\bibinfo {year} {2016})},\ \Eprint {http://arxiv.org/abs/1602.01426} {arXiv:1602.01426 [hep-lat]} \BibitemShut {NoStop}%
\bibitem [{\citenamefont {Borsanyi}\ \emph {et~al.}(2021)\citenamefont {Borsanyi} \emph {et~al.}}]{Borsanyi:2020mff}%
  \BibitemOpen
  \bibfield  {author} {\bibinfo {author} {\bibfnamefont {S.}~\bibnamefont {Borsanyi}} \emph {et~al.},\ }\href {\doibase 10.1038/s41586-021-03418-1} {\bibfield  {journal} {\bibinfo  {journal} {Nature}\ }\textbf {\bibinfo {volume} {593}},\ \bibinfo {pages} {51} (\bibinfo {year} {2021})},\ \Eprint {http://arxiv.org/abs/2002.12347} {arXiv:2002.12347 [hep-lat]} \BibitemShut {NoStop}%
\bibitem [{\citenamefont {Clarke}\ \emph {et~al.}(2024)\citenamefont {Clarke}, \citenamefont {Dimopoulos}, \citenamefont {Di~Renzo}, \citenamefont {Goswami}, \citenamefont {Schmidt}, \citenamefont {Singh},\ and\ \citenamefont {Zambello}}]{Clarke:2024ugt}%
  \BibitemOpen
  \bibfield  {author} {\bibinfo {author} {\bibfnamefont {D.~A.}\ \bibnamefont {Clarke}}, \bibinfo {author} {\bibfnamefont {P.}~\bibnamefont {Dimopoulos}}, \bibinfo {author} {\bibfnamefont {F.}~\bibnamefont {Di~Renzo}}, \bibinfo {author} {\bibfnamefont {J.}~\bibnamefont {Goswami}}, \bibinfo {author} {\bibfnamefont {C.}~\bibnamefont {Schmidt}}, \bibinfo {author} {\bibfnamefont {S.}~\bibnamefont {Singh}}, \ and\ \bibinfo {author} {\bibfnamefont {K.}~\bibnamefont {Zambello}},\ }\href@noop {} {\  (\bibinfo {year} {2024})},\ \Eprint {http://arxiv.org/abs/2405.10196} {arXiv:2405.10196 [hep-lat]} \BibitemShut {NoStop}%
\bibitem [{\citenamefont {Basar}(2024)}]{Basar:2023nkp}%
  \BibitemOpen
  \bibfield  {author} {\bibinfo {author} {\bibfnamefont {G.}~\bibnamefont {Basar}},\ }\href {\doibase 10.1103/PhysRevC.110.015203} {\bibfield  {journal} {\bibinfo  {journal} {Phys. Rev. C}\ }\textbf {\bibinfo {volume} {110}},\ \bibinfo {pages} {015203} (\bibinfo {year} {2024})},\ \Eprint {http://arxiv.org/abs/2312.06952} {arXiv:2312.06952 [hep-th]} \BibitemShut {NoStop}%
\bibitem [{\citenamefont {Cea}\ \emph {et~al.}(2016)\citenamefont {Cea}, \citenamefont {Cosmai},\ and\ \citenamefont {Papa}}]{Cea:2015cya}%
  \BibitemOpen
  \bibfield  {author} {\bibinfo {author} {\bibfnamefont {P.}~\bibnamefont {Cea}}, \bibinfo {author} {\bibfnamefont {L.}~\bibnamefont {Cosmai}}, \ and\ \bibinfo {author} {\bibfnamefont {A.}~\bibnamefont {Papa}},\ }\href {\doibase 10.1103/PhysRevD.93.014507} {\bibfield  {journal} {\bibinfo  {journal} {Phys. Rev.}\ }\textbf {\bibinfo {volume} {D93}},\ \bibinfo {pages} {014507} (\bibinfo {year} {2016})},\ \Eprint {http://arxiv.org/abs/1508.07599} {arXiv:1508.07599 [hep-lat]} \BibitemShut {NoStop}%
\bibitem [{\citenamefont {Bonati}\ \emph {et~al.}(2015)\citenamefont {Bonati}, \citenamefont {D'Elia}, \citenamefont {Mariti}, \citenamefont {Mesiti}, \citenamefont {Negro},\ and\ \citenamefont {Sanfilippo}}]{Bonati:2015bha}%
  \BibitemOpen
  \bibfield  {author} {\bibinfo {author} {\bibfnamefont {C.}~\bibnamefont {Bonati}}, \bibinfo {author} {\bibfnamefont {M.}~\bibnamefont {D'Elia}}, \bibinfo {author} {\bibfnamefont {M.}~\bibnamefont {Mariti}}, \bibinfo {author} {\bibfnamefont {M.}~\bibnamefont {Mesiti}}, \bibinfo {author} {\bibfnamefont {F.}~\bibnamefont {Negro}}, \ and\ \bibinfo {author} {\bibfnamefont {F.}~\bibnamefont {Sanfilippo}},\ }\href {\doibase 10.1103/PhysRevD.92.054503} {\bibfield  {journal} {\bibinfo  {journal} {Phys. Rev.}\ }\textbf {\bibinfo {volume} {D92}},\ \bibinfo {pages} {054503} (\bibinfo {year} {2015})},\ \Eprint {http://arxiv.org/abs/1507.03571} {arXiv:1507.03571 [hep-lat]} \BibitemShut {NoStop}%
\bibitem [{\citenamefont {Bellwied}\ \emph {et~al.}(2015)\citenamefont {Bellwied}, \citenamefont {Borsanyi}, \citenamefont {Fodor}, \citenamefont {G{\"u}nther}, \citenamefont {Katz}, \citenamefont {Ratti},\ and\ \citenamefont {Szabo}}]{Bellwied:2015rza}%
  \BibitemOpen
  \bibfield  {author} {\bibinfo {author} {\bibfnamefont {R.}~\bibnamefont {Bellwied}}, \bibinfo {author} {\bibfnamefont {S.}~\bibnamefont {Borsanyi}}, \bibinfo {author} {\bibfnamefont {Z.}~\bibnamefont {Fodor}}, \bibinfo {author} {\bibfnamefont {J.}~\bibnamefont {G{\"u}nther}}, \bibinfo {author} {\bibfnamefont {S.~D.}\ \bibnamefont {Katz}}, \bibinfo {author} {\bibfnamefont {C.}~\bibnamefont {Ratti}}, \ and\ \bibinfo {author} {\bibfnamefont {K.~K.}\ \bibnamefont {Szabo}},\ }\href {\doibase 10.1016/j.physletb.2015.11.011} {\bibfield  {journal} {\bibinfo  {journal} {Phys. Lett.}\ }\textbf {\bibinfo {volume} {B751}},\ \bibinfo {pages} {559} (\bibinfo {year} {2015})},\ \Eprint {http://arxiv.org/abs/1507.07510} {arXiv:1507.07510 [hep-lat]} \BibitemShut {NoStop}%
\bibitem [{\citenamefont {Bazavov}\ \emph {et~al.}(2019{\natexlab{b}})\citenamefont {Bazavov} \emph {et~al.}}]{Bazavov:2018mes}%
  \BibitemOpen
  \bibfield  {author} {\bibinfo {author} {\bibfnamefont {A.}~\bibnamefont {Bazavov}} \emph {et~al.},\ }\href@noop {} {\bibfield  {journal} {\bibinfo  {journal} {Physics Letters B}\ }\textbf {\bibinfo {volume} {795}},\ \bibinfo {pages} {15} (\bibinfo {year} {2019}{\natexlab{b}})},\ \Eprint {http://arxiv.org/abs/1812.08235} {arXiv:1812.08235 [hep-lat]} \BibitemShut {NoStop}%
\bibitem [{\citenamefont {Fodor}\ and\ \citenamefont {Katz}(2004)}]{Fodor:2004nz}%
  \BibitemOpen
  \bibfield  {author} {\bibinfo {author} {\bibfnamefont {Z.}~\bibnamefont {Fodor}}\ and\ \bibinfo {author} {\bibfnamefont {S.}~\bibnamefont {Katz}},\ }\href {\doibase 10.1088/1126-6708/2004/04/050} {\bibfield  {journal} {\bibinfo  {journal} {JHEP}\ }\textbf {\bibinfo {volume} {0404}},\ \bibinfo {pages} {050} (\bibinfo {year} {2004})},\ \Eprint {http://arxiv.org/abs/hep-lat/0402006} {arXiv:hep-lat/0402006 [hep-lat]} \BibitemShut {NoStop}%
\bibitem [{\citenamefont {Giordano}\ \emph {et~al.}(2020{\natexlab{a}})\citenamefont {Giordano}, \citenamefont {Kapas}, \citenamefont {Katz}, \citenamefont {Nogradi},\ and\ \citenamefont {Pasztor}}]{Giordano:2020uvk}%
  \BibitemOpen
  \bibfield  {author} {\bibinfo {author} {\bibfnamefont {M.}~\bibnamefont {Giordano}}, \bibinfo {author} {\bibfnamefont {K.}~\bibnamefont {Kapas}}, \bibinfo {author} {\bibfnamefont {S.~D.}\ \bibnamefont {Katz}}, \bibinfo {author} {\bibfnamefont {D.}~\bibnamefont {Nogradi}}, \ and\ \bibinfo {author} {\bibfnamefont {A.}~\bibnamefont {Pasztor}},\ }\href {\doibase 10.1103/PhysRevD.102.034503} {\bibfield  {journal} {\bibinfo  {journal} {Phys. Rev. D}\ }\textbf {\bibinfo {volume} {102}},\ \bibinfo {pages} {034503} (\bibinfo {year} {2020}{\natexlab{a}})},\ \Eprint {http://arxiv.org/abs/2003.04355} {arXiv:2003.04355 [hep-lat]} \BibitemShut {NoStop}%
\bibitem [{\citenamefont {Giordano}\ \emph {et~al.}(2020{\natexlab{b}})\citenamefont {Giordano}, \citenamefont {Kapas}, \citenamefont {Katz}, \citenamefont {Nogradi},\ and\ \citenamefont {Pasztor}}]{Giordano:2020roi}%
  \BibitemOpen
  \bibfield  {author} {\bibinfo {author} {\bibfnamefont {M.}~\bibnamefont {Giordano}}, \bibinfo {author} {\bibfnamefont {K.}~\bibnamefont {Kapas}}, \bibinfo {author} {\bibfnamefont {S.~D.}\ \bibnamefont {Katz}}, \bibinfo {author} {\bibfnamefont {D.}~\bibnamefont {Nogradi}}, \ and\ \bibinfo {author} {\bibfnamefont {A.}~\bibnamefont {Pasztor}},\ }\href {\doibase 10.1007/JHEP05(2020)088} {\bibfield  {journal} {\bibinfo  {journal} {JHEP}\ }\textbf {\bibinfo {volume} {05}},\ \bibinfo {pages} {088} (\bibinfo {year} {2020}{\natexlab{b}})},\ \Eprint {http://arxiv.org/abs/2004.10800} {arXiv:2004.10800 [hep-lat]} \BibitemShut {NoStop}%
\end{thebibliography}%

\end{document}